\documentclass[11pt, a4paper]{article}
\usepackage{tikz}
\usepackage[dvips]{epsfig}
\usepackage{graphics}
\usepackage{verbatim}
\usepackage{latexsym}
\usepackage{amssymb}
\usepackage{mathrsfs}
\usepackage[vlined,titlenumbered]{algorithm2e}
\newcommand\N{\mathbb{N}}
\newtheorem{Definition}{Definition}[section]
\newtheorem{Theorem}[Definition]{Theorem}
\newtheorem{Lemma}[Definition]{Lemma}

\newtheorem{Characterization}[Definition]{Characterization}
\newtheorem{Decomposition Step}[Definition]{Decomposition Step}
\newtheorem{Notation}[Definition]{Notation}
\newenvironment{Proof}{\noindent {\bf Proof:} }{{ {
$\Box$}}\vspace{\baselineskip}}
\newtheorem{Remark}[Definition]{Remark}

\newtheorem{Invariant}[Definition]{Invariant}
\newtheorem{Example}[Definition]{Example}
\newtheorem{Counterexample}[Definition]{Counterexample}

\newcommand{\ms}{minimal separator~}

\newcommand{\mt}{minimal triangulation~}

\newcommand{\msn}{minimal separator}

\title{Computing a clique tree with algorithm MLS \\ (Maximal Label Search)}
\author{Anne Berry\thanks{LIMOS UMR CNRS 6158, Ensemble Scientifique
des C\'ezeaux, F-63 173 Aubi\`ere, France, berry@isima.fr }
\and Genevi\`eve Simonet\thanks{LIRMM, 161 Rue Ada, F-34392 Montpellier,
France, genevieve.simonet@umontpellier.fr}
}
\begin{document}

\tikzset{dash/.style={dashed} } 

\maketitle

\begin{abstract}
Algorithm MLS (Maximal Label Search) is a graph search algorithm which generalizes algorithms MCS, LexBFS, LexDFS and MNS. 
On a chordal graph, MLS computes a peo 
(perfect elimination ordering) of the graph. We show how algorithm MLS can be modified
 to compute a pmo (perfect moplex ordering) as well as a 
clique tree and the minimal separators of a chordal graph. We give a necessary and
 sufficient condition on the labeling structure for the beginning of a new 
clique in the clique tree to be detected by a condition on labels. MLS is also used to compute a
 clique tree of the complement graph, and new cliques in the 
complement graph can be detected by a condition on labels for any labeling structure. 
A linear time algorithm computing a pmo and the generators of the maximal 
cliques and minimal separators w.r.t. this pmo of the complement graph is provided.
On a non-chordal graph, 
 algorithm MLSM is used to compute an atom tree of the clique minimal separator decomposition of any graph.
\end{abstract}

\section{Introduction}
Chordal graphs form an important and well-studied graph class, have many characterizations and
 properties, and are used in many applications. 
 From an algorithmic point of view, connected chordal graphs are endowed with a compact representation as a 
 \textit{clique tree}, which organizes both the maximal cliques (which are the nodes of the tree) and the minimal separators 
 (the edges): in a chordal graph, a minimal separator is the intersection of two maximal cliques, so each minimal separator is a clique (a characterization of 
 chordal graphs \cite{Dirac}). Since a connected chordal graph has at most $n$ maximal cliques, a clique tree has at most $n$ 
 nodes and less than $n$ edges, a very efficient representation of the underlying chordal graph.
 \par
 To compute a clique tree efficiently, \cite{BP} proposed an algorithm based on search algorithm MCS (Maximum Cardinality Search)
 from \cite{TY}, which numbers the vertices using labels which count the number of processed neighbors.
 MCS, as well as its famous cousin LexBFS (Lexicographic Breadth-First Search) from \cite{RTL}, were originally tailored to compute a peo (perfect elimination ordering), 
\textit{i.e.} an ordering of its vertices obtained by successively removing a simplicial vertex of
 the current graph (a vertex is simplicial if its neighborhood is a clique). 
 \par
 Recent work by Kumar and Madhaven \cite{KM} show how MCS defines the minimal separators of a chordal graph.  
 LexBFS had also been shown to scan both the maximal cliques and the minimal separators of a chordal graph 
 \cite{BBDirac}. 
 \par
This family of search algorithms 
has been recently extended by Corneil and Krueger \cite{CK}, who introduced LexDFS (Lexicographic Depth-First Search) and MNS (Maximal Neighborhood Search).
All these algorithms function on the same basic principle:
they use a labeling process to compute an ordering of the vertices
 of the input graph. All vertex labels are initialized with the same initial label. At each
 iteration of the algorithm, a yet unnumbered vertex with maximal label is chosen, and the
 labels of its yet unnumbered neighbors are increased. 
 \par
 Berry, Krueger, and Simonet \cite{MNSM}  introduced algorithm MLS (Maximal Label Search) as a generalization
 of these algorithms. 
 MLS takes as
 input a graph $G$ and a labeling structure ${\cal L}$ and computes an ordering of the vertices of $G$, which is a peo of $G$ if $G$ is
 chordal.
%  If we choose a particular labeling structure (for instance the structure 
% ${\cal L}_{LexBFS}$ implicitly used in algorithm LexBFS), we obtain an algorithm with a
%  graph $G$ as input which computes an ordering of the vertices of $G$ (algorithm LexBFS in
%  our example). 
 A condition in the definition of a labeling structure from \cite{MNSM}
 ensures that MLS computes a peo of a chordal graph.
A still more general labeling search algorithm called GLS (General Label Search) is defined
 in \cite{GLS} from a more general definition of a labeling structure, which captures
 classical categories of graph searches (general search, Breadth-First and Depth-First
 searches) in addition to the graph searches derived from MLS.
\par
The question we address in this paper is to determine in which cases MLS can be used to build a clique tree.
\par
To accomplish this, we first focus on the algorithm from \cite{BP}, Extended-MCS, which computes a clique tree of a chordal graph by computing 
the maximal
 cliques of $H$ one after the other.  
 The beginning of a new clique is detected by a
 condition on labels: as long as the label of the chosen vertex is
 strictly greater than the label of the previously chosen vertex, the current clique is
 increased, otherwise a new clique is started.
\par
Maximal cliques are strongly related to moplexes in a chordal graph.
A moplex is a clique module whose neighborhood is a \msn. In a chordal graph the union of a
 moplex and its neighborhood is a maximal clique.
Berry and Bordat \cite{BBDirac} showed that LexBFS ends on a moplex of the input graph
 whether it is chordal or not, which implies that LexBFS computes a pmo (perfect moplex
 ordering) of the input graph if it is chordal. A pmo of a graph is an ordering of its
 vertices obtained by successively removing each vertex of a simplicial moplex of the
 current graph until it is a clique.
Berry, Blair, Bordat and Simonet \cite{Extremities} proved that not only LexBFS but any instance of algorithm MLS
 with totally ordered labels computes a pmo of a chordal graph.
Berry and Pogorelcnik 
\cite{IPL} showed that MCS and LexBFS can be modified to compute the minimal separators and
 the maximal cliques of a chordal graph, using the fact that the computed ordering is a pmo
 of this graph, thus extending the result for MCS from \cite{BP}.
Xu, Li and Liang \cite{IPLLexDFS} showed that LexDFS ends on a moplex of the input graph whether it
 is chordal or not, thus extending the result for LexBFS from \cite{BBDirac}.
 They also claim that LexDFS labels detect the vertices generating the minimal separators
 and the maximal cliques of a chordal graph, as is the case for MCS or LexBFS labels, but
 this last result is erroneous, as will be seen in Section 3.
\par
In this paper, we show how a clique tree of a chordal graph $H$ can be computed from a pmo
 of $H$ by building the maximal cliques one after the other, and
how general algorithm MLS can be modified in order to compute a pmo, a clique tree and the
 minimal separators of any chordal graph for any labeling structure. We also give a
 necessary and sufficient condition on a labeling structure (which is not satisfied by 
${\cal L}_{LexDFS}$) for new cliques to be detected by a condition on labels. As MLS can be
 used to compute a peo of the complement graph of the input graph if it is chordal
 \cite{GLS}, we further modify it to compute a pmo and a clique tree of the complement
 graph, and we show that new cliques in the complement graph can be detected by a condition
 on labels for any labeling structure. We also provide a linear time algorithm computing a
 pmo and the generators of the maximal cliques and minimal separators w.r.t. this pmo of the
 complement graph.
\par
We then go on to examine what happens when the graph is not chordal.
\par
When a graph is chordal, the minimal separators are cliques.  When the graph fails to be chordal, it still may have 
clique minimal separators.  The related graph decomposition (called \textit{clique minimal separator decomposition}) 
has given rise to recent interest, see \textit{e.g. } \cite{DecompSurvey, Claw-freeTM, BraHoa2005,Leimer,Tarjan85}.  This decomposition results in a set 
of overlapping subgraphs called \textit{atoms}, characterized as the maximal connected subgraphs containing no clique separator.
\par
Recent work by Berry, Pogorelcnik and Simonet \cite{AtomTree} has shown that the atoms of a connected graph can be organized into a tree similar to a clique tree, 
called an \textit{atom tree}: 
the nodes are the atoms and the edges represent the clique minimal separators of the graph.
As is the case for a clique tree, there are at most $n$ nodes and less than $n$ edges.
\par
\cite{AtomTree} shows how an atom tree can be computed from a clique tree of a minimal triangulation (which is a 
minimal embedding of a graph into a chordal graph). It provides an algorithm based on MCS-M \cite{MCSM}, 
the triangulating counterpart of MCS, to build an atom tree.
\par
In this paper, we further address the question of using MLSM, the triangulating counterpart of MLS, 
to build this atom tree.
\par
The paper is organized as follows. Section 2 gives preliminary definitions, notations and
 known results. Section 3 explains how MLS can be modified into an algorithm computing a pmo
 and a clique tree of a chordal graph $H$, starting from a general algorithm computing a
 clique tree from a peo of $H$. In Section 4 MLS is used to compute a clique tree of the
 complement graph. Section 5 gives some extensions: the use of MLSM to compute an atom tree
 of $G$, and some counterexamples when running MLS on a non-chordal graph.

\section{Preliminaries}

 All graphs in this work are connected, undirected and finite.  A graph is denoted by $G=(V,E)$, with $|V|=n$ and $|E|=m$. 
$E(G)$ is the set of edges of $G$.
$\overline{G}$ denotes the complement of $G$.
%
%, with $\overline{m}=\vert \overline{E} \vert$.
%We say that a vertex $x$ \textit{sees} a vertex $y$ if $xy \in E$. 
%
The \textit{neighborhood} of a vertex $x$ in a graph $G$ is $N_G(x)$, 
or $N(x)$ if the context is clear; 
the \textit{closed neighborhood} of $x$ is $N[x]=N(x) \cup \{ x \}$.
The neighborhood of a subset $X$ of $V$ is $N(X) = (\cup_{x\in X}N(x)) \setminus X$, 
and its closed neighborhood is $N[X] = N(X) \cup X$.
A \textit{clique} is a set of pairwise adjacent vertices; 
%(a single vertex is also a clique); 
we say that we \textit{saturate}
a set $X$ of vertices when we add all the edges necessary to turn $X$ into %%%make $X$ 
a clique. A vertex (or a subset of $V$)
is \textit{simplicial} if its neighborhood is a clique. 
A \textit{module} is a subset $X$ of $V$ such that $\forall x \in X, N(x) \setminus X) =N(X)$. 
$G(X)$ denotes the subgraph of $G$ induced by the subset $X$ of $V$, but we will sometimes
 just denote this by $X$. 
The reader is referred to \cite{Golumbic} and \cite{BraLeSpi1999} for classical graph
 definitions and results.
\bigskip\\
\noindent
%%$$$$$$$$$$$$$$$$$$$$$$$$
\textbf{Separators.}  \\
%%$$$$$$$$$$$$$$$$$$$$$$$$
A set $S$ of vertices of a connected graph $G$ is 
a \textit{separator} if $G(V \setminus S)$ is not connected. 
A separator $S$ is an $xy$\textit{-separator} if $x$ and $y$ lie in two different connected
 components 
of $G(V \setminus S)$. 
$S$ is a \textit{minimal $xy$-separator} if $S$ is an $xy$-separator and no proper subset 
of $S$ is also an $xy$-separator. 
A separator $S$ is said to be \textit{minimal} 
if there are two vertices $x$ and $y$ such that $S$ is a minimal $xy$-separator. 
Equivalently, $S$ is a \ms if and only if $G(V\setminus S)$ has at least two connected components 
$C_1$ and $C_2$ such that $N_G(C_1)=N_G(C_2)=S$. 
%(Components such as $C_1$ and $C_2$ are called \textit{full components} of $S$ in $G$).
%A \textit{clique minimal separator} is a minimal separator which is a clique.  
%There are less than $n$ \cmss in a graph.  \\
%
%$\{x,y\}$ is a 2-pair of  $G$ if and only if $N(x) \cap N(y)$ is a minimal $xy$-separator of $G$ \cite{ArikatiRangan}. 
%
A \textit{moplex} 
%\cite{BBDirac} 
is a clique module $X$ whose neighborhood 
$N(X)$ is a \msn.
\bigskip\\
\noindent
\textbf{Chordal graphs and clique trees.}\\
%%$$$$$$$$$$$$$$$$$$$$$$$$
A graph is \textit{chordal} (or triangulated) if it contains no chordless induced cycle of length 4 or more. 
%
%A chordal graph has at most $n$ maximal cliques, and the sum of their sizes is at most $n+m$. 
%
A graph is chordal if and only if all its minimal separators are cliques \cite{Dirac}. 
It follows that for each moplex $X$ of a chordal graph $H$, $X$ is simplicial and $N[X]$ is
 a maximal clique of $H$.

A chordal graph is often represented by a \textit{clique tree}:
\begin{Definition}\label{def:clique-tree}
Let $H=(V,E)$ be a connected chordal graph. 
A {\em clique tree} of $H$ is a tree $T = (V_T,E_T)$ such that $V_T$ is the set of maximal
 cliques of $H$ and for any vertex $x$ of $H$, the set of nodes of $T$ containing $x$
 induces a subtree of $T$.
\end{Definition}
\begin{Characterization} \cite{BP}\label{carCliqueTree}
Let $H$ be a connected chordal graph, let $T$ be a clique tree of $H$, and let $S$ be a set
 of vertices of $H$; then $S$ is a \ms of $H$ if and only if 
there is an edge $K_1K_2$ of $T$ 
such that $S=K_1 \cap K_2$.
\end{Characterization}
Every chordal graph has at least one clique tree, which can be computed in linear time with
 the nodes labeled by the maximal cliques and the edges labeled by the 
minimal separators \cite{BP}.
\bigskip\\
\noindent
\textbf{Orderings, peos and pmos.}\\
An \textit{ordering} of $G$ is a one-to-one mapping from $\{1, \ldots, n\}$ to $V$. An  ordering $\alpha$ can be defined by the sequence $(\alpha(1), \ldots, \alpha(n))$.
A \textit{perfect elimination ordering (peo)} of $G$ is an ordering $\alpha = (x_1, \ldots, x_n)$ of $G$ such that for each $i \in [1,n]$, $x_i$ is a simplicial vertex of $G(\{x_i, \dots, x_n\})$.
$G$ is chordal if and only if it has a peo.
An ordering $\alpha$ of $G$ is \textit{compatible} with an ordered partition $(X_1, \ldots, X_k)$ of $V$ if for each $i$ in $[1,k-1]$, for each $u$ in $X_i$ and each $v$ in $X_{i+1}$, $\alpha ^{-1}(u) < \alpha ^{-1}(v)$.
A \textit{simple (resp. perfect) moplex partition} of $G$ is an ordered partition $(X_1, \ldots, X_k)$ of $V$ such that for each $i \in [1,k-1]$, $X_i$ is a moplex
 (resp. simplicial moplex) of $G(\cup_{i \leq j \leq n}X_j)$ and $X_k$ is a clique of $G$.
Thus a simple moplex partition of a chordal graph $H$ is a perfect moplex partition of $H$.
A \textit{perfect moplex ordering (pmo)} of $G$ is an ordering of $G$ compatible with a perfect moplex partition of $G$.
A pmo of $G$ is a peo of $G$, and $G$ is chordal if and only if it has a pmo \cite{MEO}.
\bigskip\\
\noindent
\textbf{Minimal triangulations, meos and mmos.}\\
A \textit{triangulation} of a graph $G= (V,E)$ is a chordal graph in the form $H=(V,E+F)$.
$F$ is the set of \textit{fill edges} in $H$.
%
%, $|F|$ is denoted by $f$. 
%
The triangulation is \textit{minimal} 
if for any proper subset $F^\prime$ of $F$, 
the graph $(V,E+F')$ fails to be chordal.  
Given an ordering $\alpha = (x_1, \ldots, x_n)$ of $G$, the graph $G^+_{\alpha}$
 is defined as follows:
initialize the current graph $G'$ with $G$ and 
 the set $F_{\alpha}$ with the empty set, then for each $i$ from $1$ to $n$, 
let $F_i$ be the set of edges necessary to saturate the neighborhood of $x_i$ 
in $G'$, add the edges of $F_i$ to $G'$ and to $F_{\alpha}$ and 
 remove $x_i$ from $G'$. 
$G^+_{\alpha} = (V, E+F_{\alpha})$ is a triangulation of $G$, with
$\alpha$ as a peo.
 $\alpha$ is called a \textit{minimal elimination ordering (meo)} of $G$ if there is no ordering $\beta$
 of $G$ such that $F_{\beta} \subset F_{\alpha}$.
$\alpha$ is a meo of $G$ 
if and only if $G^+_{\alpha}$ is a \mt of $G$.
A minimal triangulation of $G$ is obtained by replacing vertex $x_i$ by a moplex $X_i$ of
 $G'$ in the preceding process, which is formally defined as follows.
A \textit{minimal moplex partition} of $G$ is an ordered partition $(X_1, \ldots, X_k)$ of $V$ such that 
for each $i \in [1,k-1]$, $X_i$ is a moplex of $G_i$ and $X_k$ is a clique of $G_k$, where the graphs $G_i$ 
are defined by induction: $G_1 = G$ and for each $i \in [1,k-1]$, $G_{i+1}$ is obtained from $G_i$ by 
saturating $N_{G_i}(X_i)$ and removing $X_i$.
Thus a minimal moplex partition of a chordal graph $H$ is a perfect moplex partition of $H$.
A \textit{minimal moplex ordering (mmo)} of $G$ is an ordering  of $G$ compatible with a minimal moplex 
partition of $G$.
A mmo of $G$ is a meo of $G$ \cite{MEO}.
\bigskip\\
\noindent
\textbf{Clique Minimal separators and atom trees.}\\
The atoms of a graph $G$ are the subsets of $V$ obtained by clique minimal separator decomposition.
The reader is referred to
\cite{DecompSurvey,Leimer,Tarjan85}
for full detail on decomposition by clique separators and by clique minimal separators.
An \textit{atom} of a connected graph $G$ is a subset of $V$ inducing a connected subgraph having no clique separator, 
and being inclusion-maximal for this property.
The atoms of a chordal graph are its maximal cliques.
The atoms of a graph can be organized into an atom tree in the same way as the maximal cliques of a chordal graph are 
organized into a clique tree \cite{AtomTree}.

\begin{Definition} \cite{AtomTree} \label{def:atom-tree}
Let $G=(V,E)$ be a connected graph. 
An {\em atom tree} of $G$ is a tree $T = (V_T,E_T)$ such that $V_T$ is the set of atoms of $G$ and for any vertex $x$ of $G$, 
the set of nodes of $T$ containing $x$ induces a subtree of $T$.
\end{Definition}
\begin{Characterization} \cite{AtomTree}\label{carAtomTree}
Let $G$ be a connected graph, let $T$ be an atom tree of $G$, and let $S$ be a set of vertices of $G$; then $S$ is a 
clique \ms of $G$ if and only if 
there is an edge $A_1A_2$ of $T$ 
such that $S=A_1 \cap A_2$.
\end{Characterization}
%
%\bigskip\\
%\noindent
\textbf{Algorithms MLS and MLSM.}\\
Algorithm MLS (Maximal Label Search) generalizes the well-known algorithms MCS from \cite{TY}, LexBFS from \cite{RTL}, 
LexDFS and MNS from \cite{CK}.
Algorithm MLS has a graph $G$ and a labeling structure ${\cal L}$ as input and an ordering of $G$ as output.
It can be seen as a generic algorithm with parameter ${\cal L}$ whose instances are the algorithms ${\cal L}$-MLS for 
each labeling structure ${\cal L}$, having a graph $G$ as input and an ordering of $G$ as output. In the following 
definitions, $\N^+$ denotes the set of positive integers. 

\begin{Definition}
A {\em labeling structure} is a structure ${\cal L} = (L,\preceq,l_{0},Inc)$, where
\begin{itemize}
\item 
$L$ is a set (the set of labels), 
\item
$\preceq$ is a partial order on $L$ (which may be total or not,  with
$\prec $ denoting the corresponding strict order), 
\item
$l_{0}$ is an element of $L$ (the initial label),
\item
$Inc$ (Increase) is a mapping from $L \times \N^+$ to $L$ satisfying the following 
condition IC (Inclusion Condition):
 for any subsets $I$ and $I'$ of $\N^+$, if $I \subset I'$, then $lab_{\cal L}(I) \prec lab_{\cal L} (I')$, 
where $lab_{\cal L} (I) = Inc( \ldots (Inc(l_0,i_1), \ldots),i_k)$,
where $I = \{i_1,i_2, \ldots,i_k\}$, with $i_1 > \cdots > i_k$.
\end{itemize}
\end{Definition}

For each $X \in \{MCS, LexBFS, LexDFS, MNS\}$, algorithm X
is the instance ${\cal L}_X$-MLS of algorithm MLS, where ${\cal L}_X$ is the labeling structure $(L,\preceq,l_{0},Inc)$ 
defined as follows.

${\cal L}_{MCS}$:
$L = \N^+ \cup \{0\}$,
$\preceq$ is $\leq$ (a total order), 
$l_{0} = 0$, 
$Inc(l,i) = l + 1$.

${\cal L}_{LexBFS}$:
$L$ is the set of  lists of elements of $\N^+$, 
$\preceq$ is the usual lexicographic order (a total order), 
$l_{0}$ is the empty list, 
$Inc(l,i)$ is obtained from $l$ by appending $i$ to the end of the list.

${\cal L}_{LexDFS}$:
$L$ is the set of  lists of elements of $\N^+$, 
$\preceq$ is the lexicographic order where the order on $\N^+$ is the reverse order of the usual one (a total order), 
$l_{0}$ is the empty list, 
$Inc(l,i)$ is obtained from $l$ by prepending $i$ to the beginning of the list.

${\cal L}_{MNS}$:
$L$ is the power set of $\N^+$, 
$\preceq$ is $\subseteq$ (not a total order), 
$l_{0} = \emptyset$, 
$Inc(l,i) = l \cup \{i\}$.

\begin{algorithm}[H]
\SetKwInOut{Input}{input}
\SetKwInOut{Output}{output}
\textbf{Algorithm MLS} 
\BlankLine
\Input{a connected graph $H$ and a labeling structure ${\cal L} = (L, \preceq,l_{0},Inc)$}
\Output{an ordering $\alpha = (x_1, \ldots, x_n)$ of $H$, which is a peo of $H$ if $H$ is chordal}
\BlankLine
{ % begin
$V' \leftarrow \emptyset$; Initialize all labels as $l_{0}$\;
\For{ $i=n$  \textbf{downto} $1$}
{ % for
Choose a vertex $x$ in $V \setminus V'$ whose label is maximal\;
$x_i \leftarrow x$\;  
\ForEach {$y$ in $N_{H}(x) \setminus V'$}
{
$label(y) \leftarrow Inc(label(y),i)$\;
}
$V' \leftarrow V' \cup \{x\}$\;
} % for
} % begin
\end{algorithm}

\begin{algorithm}[H]
\SetKwInOut{Input}{input}
\SetKwInOut{Output}{output}
\textbf{Algorithm MLSM} 
\BlankLine
\Input{a connected graph $G$ and a labeling structure ${\cal L} = (L, \preceq,l_{0},Inc)$}
\Output{a meo $\alpha = (x_1, \ldots, x_n)$ of $G$ and the associated minimal triangulation $H = G^+_{\alpha}$}
\BlankLine
{ % begin
$H \leftarrow G$; $V' \leftarrow \emptyset$; Initialize all labels as $l_{0}$\;
\For{ $i=n$  \textbf{downto} $1$}
{ % for1
Choose a vertex $x$ in $V \setminus V'$ whose label is maximal\; 
$Y \leftarrow \emptyset$\;
\ForEach {$y$ in $V \setminus (V' \cup \{x\})$}
{ % for2
\If {there is a path $\mu$ of length $\geq 1$ in $G(V \setminus V')$ between $x$ and $y$ such that for each 
internal vertex $z$ of $\mu$ $label(z) \prec label(y)$} 
{ % if
$Y \leftarrow Y + \{y\}$; $E(H) \leftarrow E(H) \cup \{xy\}$\;
} % if
} % for2
\ForEach {$y$ in $Y$}
{
$label(y) \leftarrow Inc(label(y),i)$\;
}
$V' \leftarrow V' \cup \{x\}$;
} % for1
} % begin
\end{algorithm}

Algorithms LEX M from \cite{RTL} and MCS-M from \cite{MCSM} are the instances ${\cal L}_{LexBFS}$-MLSM and ${\cal L}_{MCS}$-MLSM 
respectively of algorithm MLSM.

\begin{Notation}
Let $G$ be a graph, let $\alpha$ be an ordering of $G$, let $i \in [1,n]$ and let $y \in V$.

- $N_G^{\alpha ,i}(y) = \{z \in N_G(y)~|~\alpha^{-1}(z) > i\}$,

- $N_G^{\alpha +}(y) = N_G^{\alpha , \alpha^{-1}(y)}(y) = \{z \in N_G(y)~|~\alpha^{-1}(z) > \alpha^{-1}(y) \}$,

- $N_G^{\alpha +}[y] = N_G^{\alpha +}(y) \cup \{y\} = \{z \in N_G[y]~|~\alpha^{-1}(z) \geq \alpha^{-1}(y) \}$,

- a {\em generator} of a minimal separator $S$ (resp. maximal clique $K$) of $G$ w.r.t. $\alpha$ is a vertex $x$ of $G$ 
such that $S = N_G^{\alpha +}(x)$ (resp. $K = N_G^{\alpha +}[x]$).
\end{Notation}

If $\alpha$ is a peo of a chordal graph $H$ then each minimal separator of $H$ has at least 
one generator w.r.t. $\alpha$ \cite{Rose} 
and each maximal clique of $H$ clearly has exactly one generator w.r.t. $\alpha$, which is the vertex $x$ of $K$ 
with minimum $\alpha ^{-1}(x)$.

\begin{Lemma} \label{lemIC}
In an execution of MLS, 
for each $i$ in $[1,n-1]$ and each $y,z$ in $V$ such that $\alpha^{-1}(y) \leq i+1$ and $\alpha^{-1}(z) \leq i+1$, 
at the beginning of iteration $i$ (choosing vertex $x_i$) of the \textbf{for} loop,

i) If $N_G^{\alpha ,i}(y) \subset N_G^{\alpha ,i}(z)$ then $label(y) \prec label(z)$,

ii) If $N_G^{\alpha ,i}(y) \subseteq N_G^{\alpha ,i}(z)$ then $label(y) \preceq label(z)$.

\end{Lemma}

\begin{Proof}
i) If $N_G^{\alpha ,i}(y) \subset N_G^{\alpha ,i}(z)$ then 
$\alpha^{-1}(N_G^{\alpha ,i}(y)) \subset \alpha^{-1}(N_G^{\alpha ,i}(z))$, and therefore by condition IC
$label(y) = lab_{\cal L}(\alpha^{-1}(N_G^{\alpha ,i}(y))) \prec lab_{\cal L}(\alpha^{-1}(N_G^{\alpha ,i}(z))) = label(z)$.
The proof of ii) is similar.
\end{Proof}

\section{MLS and clique trees}

In this section, we show how general algorithm MLS can be used to compute a clique tree and the minimal separators of a 
chordal graph, thus generalizing the results given in \cite{BP} for MCS and in \cite{IPL} for LexBFS.
We will do this by applying successive modifications on an algorithm computing a clique tree from a peo of a chordal graph.

\subsection{Clique tree from a peo}

Algorithm CliqueTree from Spinrad (\cite{Spinrad2003} p. 258) computes a clique tree of a connected chordal graph from an 
arbitrary peo of this graph.

\begin{algorithm}[H]
\SetKwInOut{Input}{input}
\SetKwInOut{Output}{output}
\textbf{Algorithm CliqueTree} 
\BlankLine
\Input{a connected chordal graph $H$ and a peo $\alpha = (x_1, \ldots, x_n)$ of $H$}
\Output{a clique tree $T$ and the set $Sep$ of minimal separators of $H$}
\BlankLine
{ % begin
$V' \leftarrow \emptyset$; $s \leftarrow 1$; $K_1 \leftarrow \emptyset$; $E \leftarrow \emptyset$; $Sep \leftarrow \emptyset$\; 
\For{ $i=n$  \textbf{downto} $1$}
{ % for
$x \leftarrow x_i$; $S \leftarrow N_H(x) \cap V'$;
// $S = N_H^{\alpha+}(x)$ \\
\If {$i = n$} 
{
$p \leftarrow 1$\;
}
\Else
{
$k \leftarrow min\{j, \alpha(j) \in S\}$;  // $S \neq \emptyset$ because $H$ is connected,  $\alpha$ is a peo of $H$ and $i<n$ \\
$p \leftarrow clique(\alpha(k)) $\; 
}
\If {$K_p = S$} 
{
$clique(x) \leftarrow p$\;
}
\Else
{
$s \leftarrow s+ 1; K_s \leftarrow S$;  // start a new clique \\
$E \leftarrow E \cup \{(p,s)\} $;
$Sep \leftarrow Sep \cup \{S\}$; // $S = K_p \cap K_s$ \\
$clique(x) \leftarrow s$\;
}
$K_{clique(x)} \leftarrow K_{clique(x)} \cup \{x\}$; // increase clique \\
$V' \leftarrow V' \cup \{x\}$;
} % for
$T \leftarrow (\{K_1, \dots,, K_s\},  \{K_p K_q,~(p,q) \in E\})$;
} % begin
\end{algorithm}

\begin{Theorem} \label{thCT}
\cite{Spinrad2003}
Algorithm CliqueTree computes a clique tree and the set of minimal separators of a connected chordal graph $H$ from a peo of $H$ 
in linear time.
\end{Theorem}

Note that the algorithm from \cite{Spinrad2003} only computes a clique tree of the input graph. 
By Characterization~\ref{carCliqueTree} Algorithm CliqueTree correctly computes the set of minimal separators, and it does 
it in linear time using a Search/Insert structure for $Sep$ allowing to check the presence and insert a set $S$ in $O(|S|)$ time.
The proof given in \cite{Spinrad2003} implicitly uses following Invariant~\ref{invCT} using itself  Lemma~\ref{lemCliqueTreeIncrement} 
below, which are explicitly stated and proved here since they will be used later in this paper.

\begin{Invariant} \label{invCT}
The following proposition is an invariant of the \textbf{for} loop of Algorithm CliqueTree:
\begin{itemize}
\item[a)] $(\{K_1, \dots, K_s\},  \{K_p K_q,~(p,q) \in E\})$ is a clique tree of $H(V')$,
\item[b)] $Sep$ is the set of minimal separators of  $H(V')$,
\item[c)] $\forall y \in V'$, $clique(y) \in [1,s]$ and $N_H^{\alpha+}[y]   \subseteq K_{clique(y)}$. 
\end{itemize}
\end{Invariant}

\begin{Proof} 
The proposition clearly holds at the initialization step of the \textbf{for} loop. Let us show that it is preserved 
by each iteration  of this loop. It is clearly preserved by iteration $n$, \textit{i.e.} the iteration where $i=n$. Let us show 
that it is preserved by iteration $i$, with $1 \leq i<n$. 
Let a1) (resp. b1), c1)) denote item a) (resp. b), c)) at the beginning of iteration $i$ (which is supposed to be true) 
and let $x$, $V'$, $S$, $s$, $k$ and $p$ be the values of these variable at the end of iteration $i$. 
As $\alpha$ is a peo of $H$, $x$ is a simplicial vertex of $H(V')$, so $S$ is a clique of $H$. It follows by definition 
of $k$ that $S \subseteq N_H^{\alpha+}[\alpha(k)]$, hence by c1) $S \subseteq K_p$ with $p \in [1,s]$. Thus, by a1) $K_p$ 
is a maximal clique of $H(V' \setminus \{x\})$ containing $S$. It follows from Lemma~\ref{lemCliqueTreeIncrement} 
and Characterization~\ref{carCliqueTree}
that a) and b) are preserved. It remains to show that c) is preserved. It is the case since 
for each $y \in V' \setminus \{x\}$ $clique(y)$ and $N_H^{\alpha+}[y]$ are unchanged whereas $s$ 
and $K _{clique(y)}$ can only become bigger at iteration $i$, and for $y=x$, $clique(x)$ is either equal 
to $p$ or to $s$ with $p\in [1,s]$, 
and $N_H^{\alpha+}[x] = S \cup \{x\} = K_{clique(x)}$.
\end{Proof} 

\begin{Lemma} \label{lemCliqueTreeIncrement}
Let $H$ be a connected chordal graph,
let $x$ be a simplicial vertex of $H$,
let $H' = H(V \setminus \{x\})$,
let $T'$ be a clique tree of $H'$,
let $K$ be  a node of $T'$ containing $N_H(x)$,
let $K' = N_H[x]$,
and let $T$ be the tree obtained from $T'$ by replacing node $K$ by $K'$ (with the same neighbors in $T$ as in $T'$) if $K = N_H(x)$ 
and by adding node $K'$ and edge $KK'$ 
otherwise.
Then $T$ is a clique tree of $H$.
\end{Lemma}

\begin{Proof} 
Let us show that the nodes of $T$ are the maximal cliques of $H$. As $x$ is simplicial in $H$, $K'$ is the unique maximal clique 
of $H$ containing $x$. Each maximal clique of $H$ different from $K'$ is a maximal clique of $H'$, and each maximal clique of $H'$ 
 different from $N_H(x)$ is a maximal clique of $H$. It follows that the nodes of $T$ are the maximal cliques of $H$. It remains to 
show that for each vertex $y$ of $H$, the subgraph $T_y$ of $T$ induced by the set of nodes of $T$ containing $y$ is connected. 
If $y=x$ then $T_y$ is reduced to node $K'$ otherwise $T_y$ is either equal to $T'_y$ or obtained from $T'_y$ by replacing node $K$ 
by node $K'$ or by nodes $K$ and $K'$ and edge $KK'$. Hence $T_y$ is connected.
\end{Proof} 

\subsection{Clique tree from a pmo}

According to Invariant~\ref{invCT}, in an execution of Algorithm CliqueTree  the cliques $K_1, \dots, K_s$ are the maximal cliques 
of $H(V')$, and therefore cliques of $H$ which are not necessarily maximal in $H$.  Some peos build the maximal cliques of $H$ one 
after the other, that is, at each time in an execution of the algorithm, each clique $K_j$  different from $K_s$ is a maximal 
clique of $H$, and a vertex is added to $K_s$ at each iteration of the \textbf{for} loop until $K_s$  is a maximal clique of $H$ 
and $s$ is incremented to start a new  maximal clique of $H$.
If $\alpha = (x_1, \ldots, x_n)$ is such a peo, at the beginning of iteration $i<n$ the current clique $K_s$ is equal 
to $N_H^{\alpha+}[x_{i+1}]$ and 
if $K_s$ is not a maximal clique of $H$ then $K_s = S = N_H^{\alpha+}(x_i)$ and $x_i$ is added to $K_s$ at iteration $i$.

\begin{Definition}
A {\em MCComp (Maximal Clique Completing) peo} of a connected chordal graph $H$ is a peo $\alpha = (x_1, \ldots, x_n)$ 
of $H$
such that for each $i \in [1,n-1]$, $N_H^{\alpha+}[x_{i+1}]$ is a maximal clique of $H$ or equal to $N_H^{\alpha+}(x_i)$.
\end{Definition} 

\begin{figure}
\begin{center}
\begin{tikzpicture} 

%\begin{scope}
\coordinate (1) at (0,2) ;
\draw (1) node {$\bullet$}
               node  [above] {$a$};

\coordinate (2) at (0,0) ;
\draw  (2) node {$\bullet$}
               node [below] {$b$};
               
 \coordinate (3) at (5,2) ;
\draw  (3) node {$\bullet$}
               node  [above] {$c$};
               
\coordinate (4) at (5,0) ;
\draw  (4) node {$\bullet$}
               node  [below] {$d$};
                
\coordinate (5) at (4,1) ;
\draw  (5) node {$\bullet$}
               node  [below] {$e$};          
                
\coordinate (6) at (1,1) ;
\draw  (6) node {$\bullet$}
               node  [below] {$f$};  
              
\draw (6) -- (1) -- (2) -- (6) -- (5) -- (3) -- (4) -- (5) ;

\draw  (0.4,1) node {$K$};
\draw  (4.6,1) node {$K'$};
%\end{scope}      

\end{tikzpicture}
\end{center} 

\caption{A chordal graph $H$.}
\label{figchordal}

\end{figure}
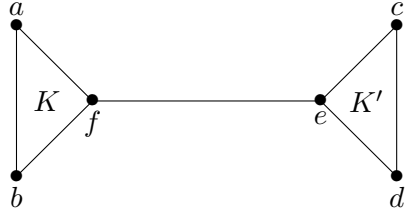

\begin{Example}
Let $H$ be the graph shown in Figure~\ref{figchordal}, whose maximal cliques are $K = \{a,b,f\}$, $K' = \{c,d,e\}$ and $\{e,f\}$,
and let $\alpha = (a,b,c,d,e,f)$. An execution of CliqueTree on $H$ and $\alpha$ successively completes the maximal cliques $\{e,f\}$, 
$K'$ and $K$, and we easily check that $\alpha$ is a MCComp peo of $H$.
Now let $\beta = (a,c,d,b,e,f)$.
An execution of CliqueTree on $H$ and $\beta$ successively completes $\{e,f\}$, starts $K$, starts and completes $K'$, 
and finally completes $K$. 
$\beta$ is not a MCComp peo of $H$ since for $i = 3$, $N_H^{\beta+}[x_{i+1}]$ is neither a maximal clique of $H$ nor equal 
to $N_H^{\beta+}(x_i)$ ($N_H^{\beta+}[x_{i+1}] = N_H^{\beta+}[b] = \{b,f\}$ and $N_H^{\beta+}(x_i) = N_H^{\beta+}(d) = \{e\}$).
\end{Example} 

%%%%%%%%%%%%%%%%%%%%%%%%%%%%%%%

%\begin{figure}
%\centerline{\includegraphics[width=7cm]{fig1}}
%\caption{A chordal graph $H$.}
%\label{figchordal}
%\end{figure} 

%%%%%%%%%%%%%%%%%%%%%%%%%%%%%%%

Using a MCComp peo instead of an arbitrary peo, Algorithm CliqueTree can be simplified into Algorithm MCComp-CliqueTree 
containing the blocks \textbf{InitCT}, \textbf{StartClique}, \textbf{IncreaseClique}, and \textbf{DefineCT} which will be used in 
further algorithms in this paper.

\begin{algorithm}[H]
\textbf{InitCT} \\
%\BlankLine 
{
$V' \leftarrow \emptyset$; $s \leftarrow 1$; $K_1 \leftarrow \emptyset$; $E \leftarrow \emptyset$; $Sep \leftarrow \emptyset$; 
}
\end{algorithm}

\begin{algorithm}[H]
\textbf{StartClique} \\
%\BlankLine 
{
$s \leftarrow s+ 1; K_s \leftarrow S$\; 
$k \leftarrow min\{j, \alpha(j) \in S\}$;
$p \leftarrow clique(\alpha(k)) $; 
$E \leftarrow E \cup \{(p,s)\} $;
$Sep\leftarrow Sep \cup \{S\}$;
}
\end{algorithm}

\begin{algorithm}[H]
%\SetKwInOut{Input}{input}
\textbf{IncreaseClique} \\
%\BlankLine
%\Input{an integer $q$}
{
$K_s \leftarrow K_s \cup \{x\}$; $clique(x) \leftarrow s$; 
}
\end{algorithm}

\begin{algorithm}[H]
\textbf{DefineCT} \\
%\BlankLine 
{
$T \leftarrow (\{K_1, \dots, K_s\},  \{K_p K_q,~(p,q) \in E\})$;
}
\end{algorithm}

\begin{algorithm}[H]
\SetKwInOut{Input}{input}
\SetKwInOut{Output}{output}
\textbf{Algorithm MCComp-CliqueTree} 
\BlankLine
\Input{a connected chordal graph $H$ and a MCComp peo $\alpha = (x_1, \ldots, x_n)$ of $H$}
\Output{a clique tree $T$ and the set $Sep$ of minimal separators of $H$}
\BlankLine
{ % begin
\textbf{InitCT}\;
\For{ $i=n$  \textbf{downto} $1$}
{ % for
$x \leftarrow x_i$;  $S \leftarrow N_H(x) \cap V'$;
// $S = N_H^{\alpha+}(x)$ \\
\If {$K_s \neq S$} 
{
\textbf{StartClique}\;
}
\textbf{IncreaseClique}\;
$V' \leftarrow V' \cup \{x\}$;
} % for
\textbf{DefineCT};
} % begin
\end{algorithm}

\begin{Theorem} \label{thMCT}
Algorithm MCComp-CliqueTree computes a clique tree and the set of minimal separators of a connected chordal graph $H$ from a 
MCComp peo of $H$ in linear time.
\end{Theorem}

\begin{Proof} 
The proof of complexity is similar to that of Theorem~\ref{thCT}, and correctness of the algorithm follows from, 
Invariant~\ref{invMCT}.
\end{Proof} 

\begin{Invariant} \label{invMCT}
The following proposition is an invariant of the \textbf{for} loop of Algorithm MCComp-CliqueTree:
\begin{itemize}
\item[a)] b)~ c)  (as in Invariant~\ref{invCT})
\item[d)] $\forall j \in [1,s-1]$ $K_j$ is a maximal clique of $H$,
\item[e)] $K_s =N_H^{\alpha+}[x_{i}]$.
\end{itemize}
\end{Invariant}

\begin{Proof} 
The proposition  clearly holds at the initialization step (except for e) which is undefined). Let us shows that it is preserved 
at iteration $i$, with $1 \leq i \leq n$. 
Let a1) (resp. b1), \ldots, e1)) denote item a) (resp. b), \ldots, e)) at the beginning of iteration $i$ (which is supposed to be true) 
and let $s$ be the value of this variable at the beginning of iteration $i$. 
We prove that a), b) and c) are preserved as in the proof of Invariant~\ref{invCT}  except that we have moreover to show that 
if $K_s \neq S$ then $K_p \neq S$.
It is evident if $p = s$, otherwise it follows from the fact that  $K_p$ is a maximal clique of $H$ by d1) whereas $S$ is not 
since it is a strict subset of the clique $S \cup \{x\}$.
Hence a), b) and c) are preserved. 
Let us show that d) is preserved. We only have to check that in case $s$ is incremented (to $s+1$), $K_{s}$ is a maximal clique 
of $H$. 
As $s$ is incremented at iteration $i$ $K_s \neq S$, so $i <n$ and therefore by e1) $K_s = N_H^{\alpha+}[x_{i+1}]$. 
As $K_s \neq S$ with $S = N_H^{\alpha+}(x_i)$ and $\alpha$ is a MCComp peo of $H$, it follows that $K_s$ is a maximal clique of $H$. 
Thus d) is preserved, and e) obviously holds at the end of iteration $i$.
\end {Proof} 

\begin{Characterization} \label{carMoplex}
An ordering $\alpha$ of  a connected chordal graph $H$ is a MCComp peo of $H$ if and only if it is a pmo of $H$. 
\end{Characterization}

\begin{Proof} 
We prove this by induction on $n  = |V|$. It trivially holds if 
$n=1$. We suppose that it holds if $|V| <n$. Let us show  that it holds if $|V|= n$. Let $\alpha = (x_1, \ldots, x_n)$ be an ordering 
of $H$ and let $H' = H(V \setminus \{x_1\})$. \\
$\Rightarrow$: We suppose that $\alpha$ is a MCComp peo of $H$. Let us show that it is a pmo of $H$. 
$(x_2, \ldots, x_n)$ is a MCComp peo of $H'$ so by induction hypothesis it is a pmo of $H'$ compatible with a perfect moplex partition 
of $H'$, say $(X_1, \ldots, X_k)$.
If $N^{\alpha +}_H(x_1) = N^{\alpha +}_H[x_2]$ then $(\{x_1\} \cup X_1,  \ldots, X_k)$ is a perfect moplex partition of $H$.
Otherwise, in an execution of Algorithm MCComp-CliqueTree on $H$ and $\alpha$, $K_s \neq S$ at iteration 1 
since $S =  N^{\alpha +}_H(x_1)$ and $K_s = N^{\alpha +}_H[x_2]$ by Invariant~\ref{invMCT} e), so $S$ is a minimal 
separator of $H$, which makes $\{x_1\}$ a moplex of $H$ and $(\{x_1\}, X_1,  \ldots, X_k)$  a perfect moplex partition of $H$. 
Hence $\alpha$ is a pmo of $H$. \\
$\Leftarrow$: We suppose that $\alpha$ is a pmo of $H$ compatible with perfect moplex partition $(X_1,  \ldots, X_k)$. 
Let us show that it is a MCComp peo of $H$. 
As it is a pmo of $H$ it is a peo of $H$.
$(x_2, \ldots, x_n)$ is a pmo of $H'$ (compatible with perfect moplex partition  $(X_2,  \ldots, X_k)$ if $X_1 = \{x_1\}$ 
and $(X_1 \setminus  \{x_1\}, \ldots, X_k)$ otherwise)
so by induction hypothesis it is a MCComp peo of $H'$.
Hence for each $i$ from 2 to $n-1$, $N^{\alpha +}_H[x_{i+1}]$ is a maximal clique of $H'$ or equal to $N^{\alpha +}_H(x_i)$.
It is sufficient to show that for each $i$ from 2 to $n-1$, if $N^{\alpha +}_H[x_{i+1}]$ is a maximal clique of $H'$ then it is a 
maximal clique of $H$, and that $N^{\alpha +}_H[x_2]$ is a maximal clique of $H$ or equal to $N^{\alpha +}_H(x_1)$. \\
\textit{First case:} $X_1 = \{x_1\}$ \\
As $N_H(x_1)$ is a minimal separator of $H$, by Characterization~\ref{carCliqueTree}
it is equal to the intersection of two maximal cliques of $H$, and therefore is not a maximal clique of $H'$. It follows 
by Lemma~\ref{lemCliqueTreeIncrement} that each maximal clique of $H'$ is a maximal clique of $H$. 
Moreover $N^{\alpha +}_H[x_2] = N_{H'}[X_2]$, which is a maximal clique of $H'$ and therefore of $H$. \\
\textit{Second case:} $X_1 \neq \{x_1\}$ \\
In that case $N^{\alpha +}_H(x_1) = N^{\alpha +}_H[x_2]$. By Lemma~\ref{lemCliqueTreeIncrement} each maximal clique of $H'$ 
different from $N_H(x_1)$ is a maximal clique of $H$. It follows that for each $i$ from 2 to $n-1$, if $N^{\alpha +}_H[x_{i+1}]$ 
is a maximal clique of $H'$ then it is a maximal clique of $H$ since it does not contain $x_2$ whereas $N_H(x_1)$ contains $x_2$.
\end{Proof} 

\subsection{Clique tree using MLS}

Algorithms MCS, LexBFS and LexDFS, and more generally algorithm ${\cal L}$-MLS for any labeling structure ${\cal L}$ for which the 
order on labels is total, compute a pmo of a connected chordal graph \cite{Extremities}. 
Note that the definition of a labeling structure given in \cite{Extremities} is less general than the definition given in this paper, 
but the proof of this result still holds here.
 We define Algorithm Moplex-MLS, which computes a pmo of a chordal graph, whether the order on labels is total or not, by adding a 
 tie-breaking rule when choosing a vertex with a maximal label when the ordering fails to be total. 

\begin{algorithm}[H]
\SetKwInOut{Input}{input}
\SetKwInOut{Output}{output}
\textbf{Algorithm Moplex-MLS} 
\BlankLine
\Input{a connected chordal graph $H$ and a labeling structure ${\cal L} = (L, \preceq,l_{0},Inc)$}
\Output{a pmo $\alpha = (x_1, \ldots, x_n)$ of $H$}
\BlankLine
{ % begin
$V' \leftarrow \emptyset$; Initialize all labels as $l_{0}$; $prev$-$max$-$label \leftarrow l_0$\;
\For{ $i=n$  \textbf{downto} $1$}
{ % for
Choose a vertex $x$ in $V \setminus V'$ whose label is maximal, \\
and if possible strictly greater than $prev$-$max$-$label$\;
$x_i \leftarrow x$\; 
\ForEach {$y$ in $N_{H}(x) \setminus V'$}
{
$label(y) \leftarrow Inc(label(y),i)$\;
}
$V' \leftarrow V' \cup \{x\}$; $prev$-$max$-$label \leftarrow label(x)$\;
} % for
} % begin
\end{algorithm}

\begin{Theorem} \label{thMMLS}
Algorithm Moplex-MLS computes a pmo of a connected chordal graph.
\end{Theorem}

To prove  Theorem~\ref{thMMLS} we will use the following Lemma.

\begin{Lemma} \label{lemMMLS}
In an execution of Moplex-MLS, 
for each $i$ in $[1,n-1]$ and each $y$ in $V$ such that $\alpha^{-1}(y) \leq i$, 
at the beginning of iteration $i$,
\begin{itemize}
\item[a)] If $N_H^{\alpha+}[x_{i+1}] \subseteq N_H^{\alpha ,i}(y)$ then  $prev$-$max$-$label \prec label(y)$.
\item[b)] If  $prev$-$max$-$label \prec label(y)$ then $\{x_{i+1}\} \subseteq N_H^{\alpha ,i}(y) \subseteq N_H^{\alpha+}[x_{i+1}]$.
\end{itemize}
\end{Lemma}

\begin{Proof} 
a) If $N_H^{\alpha+}[x_{i+1}] \subseteq N_H^{\alpha ,i}(y)$ then $N_H^{\alpha ,i}(x_{i+1}) \subset N_H^{\alpha ,i}(y)$, so 
by Lemma~\ref{lemIC}
 $prev$-$max$-$label = label(x_{i+1}) \prec label(y)$. \\
b) We suppose that $prev$-$max$-$label \prec label(y)$. As the label of $x_{i+1}$ is maximal at the beginning of 
iteration $i+1$, $label(y)$ has been increased during iteration $i+1$, so $y$ is a neighbor of $x_{i+1}$ in $H$. As $\alpha$ is 
a MLS ordering of $H$, it is a peo of $H$, so $N_H^{\alpha+}(y)$ is a clique containing $x_{i+1}$, 
and therefore $N_H^{\alpha ,i}(y) \subseteq N_H^{\alpha+}[x_{i+1}]$. 
\end{Proof} 

\begin{Proof} (of Theorem~\ref{thMMLS})
Let $\alpha = (x_1, \ldots, x_n)$ be the ordering computed by an execution of Moplex-MLS on input $H$ and ${\cal L}$. 
Let us show that it is a pmo of $H$.
By Characterization~\ref{carMoplex}
it is sufficient to show that $\alpha$ is a MCComp peo of $H$. 
As $\alpha$ is a MLS ordering of $H$, it is a peo of $H$.
Let $i \in [1,n-1]$. We suppose that $N_H^{\alpha+}[x_{i+1}]$ is not a maximal clique of $H$. Let us show that it is 
equal to $N_H^{\alpha+}(x_{i})$. 
As $N_H^{\alpha+}[x_{i+1}]$  is not a maximal clique of $H$, there is a vertex $y$ such that $\alpha^{-1}(y) \leq i$ 
and $N_H^{\alpha+}[x_{i+1}] \subseteq N_H^{\alpha ,i}(y)$, and therefore 
by Lemma~\ref{lemMMLS} a) $prev$-$max$-$label \prec  label(y)$ at the beginning of iteration $i$ in this execution. 
It follows by the condition on the choice of $x$
 that $prev$-$max$-$label \prec label(x_{i})$, and therefore by Lemma~\ref{lemMMLS} b)
 $N_H^{\alpha+}(x_{i}) \subseteq N_H^{\alpha+}[x_{i+1}]$. 
It is impossible that $N_H^{\alpha+}(x_{i}) \subset N_H^{\alpha+}[x_{i+1}]$ since in that case
$N_H^{\alpha ,i}(x_{i}) \subset N_H^{\alpha ,i}(y)$, so
 by Lemma~\ref{lemIC} at the beginning of iteration $i$ $label(x_{i}) \prec label(y)$ and $x_{i}$ would not be a vertex with maximal label. Hence $N_H^{\alpha+}(x_{i}) = N_H^{\alpha+}[x_{i+1}]$.
\end{Proof} 

If ${\cal L}$ is a labeling structure with a total order on labels, condition "if possible strictly greater than $prev$-$max$-$label$" 
is useless, so Algorithm Moplex-${\cal L}$-MLS is actually identical to ${\cal L}$-MLS. Thus we refind the result 
from \cite{Extremities} that if the order on labels is total then each ${\cal L}$-MLS ordering of a chordal graph is 
a pmo of this graph, with a more general definition of a labeling structure and an alternative (simpler) proof. 
\par
Combining Algorithms MCComp-CliqueTree and Moplex-MLS, we define Algorithm MLS-CliqueTree computing both a pmo and a clique tree 
of a chordal graph.

\begin{algorithm}[H]
\SetKwInOut{Input}{input}
\SetKwInOut{Output}{output}
\textbf{Algorithm MLS-CliqueTree} 
\BlankLine
\Input{a connected chordal graph $H$ and a labeling structure ${\cal L} = (L, \preceq,l_{0},Inc)$}
\Output{ a pmo $\alpha = (x_1, \ldots, x_n)$,
a clique tree $T$ and the set $Sep$ of minimal separators of $H$}
\BlankLine
{ % begin
\textbf{InitCT}\; 
Initialize all labels as $l_{0}$; $prev$-$max$-$label \leftarrow l_0$\;
\For{ $i=n$  \textbf{downto} $1$}
{ % for
Choose a vertex $x$ in $V \setminus V'$ whose label is maximal, \\
and if possible strictly greater than $prev$-$max$-$label$\;
$x_i \leftarrow x$;  $S \leftarrow N_H(x) \cap V'$;
// $S = N_H^{\alpha+}(x)$ \\
\If {$K_s \neq S$} 
{
\textbf{StartClique}\;
}
\textbf{IncreaseClique}\;
\ForEach {$y$ in $N_{H}(x) \setminus V'$}
{
$label(y) \leftarrow Inc(label(y),i)$\;
}
$V' \leftarrow V' \cup \{x\}$; $prev$-$max$-$label \leftarrow label(x)$\;
} % for
\textbf{DefineCT};
} % begin
\end{algorithm}

Correctness of Algorithm MLS-CliqueTree immediately follows from the correctness of Algorithms Moplex-MLS and MCComp-CliqueTree 
and from Characterization~\ref{carMoplex}.

\begin{Example}
Consider an execution of Algorithm MLS-CliqueTree on the graph $H$ shown in Figure~\ref{figchordal} and labeling structure ${\cal L}_X$ 
with $X \in \{MCS, LexBFS, LexDFS, MNS\}$ choosing vertices $f$ then $e$ first (and therefore completing the maximal 
clique $\{e,f\}$ first).
Then the execution successively completes $K$ then $K'$ if $X = LexBFS$,
$K'$ then $K$ if $X = LexDFS$, and either $K$ then $K'$ or $K'$ then $K$ otherwise.
Removing condition " if possible strictly greater than $prev$-$max$-$label$" has no effect if $X \neq MNS$, but if $X = MNS$ it would 
allow the execution to choose alternatively a vertex of $K$ and a vertex of $K'$, as the labels of the vertices of $K$ are incomparable 
to the labels of the vertices of $K'$.
\end{Example} 

 Algorithm MLS-CliqueTree generalizes algorithm Extended-MCS from \cite{BP} and its extension to LexBFS from \cite{IPL}, except that in 
these algorithms, the condition "$K_s \neq S$" is replaced by a direct 
condition on labels: "$label(x) \preceq prev$-$max$-$label$". We define a necessary and sufficient condition on a 
labeling structure ${\cal L}$ for the replacement of "$K_s \neq S$" by "$prev$-$max$-$label \not \prec label(x)$" 
(which becomes "$label(x) \preceq prev$-$max$-$label$" if $\preceq$ is a total order) to be possible.

\begin{Definition}
Let ${\cal L}=(L,\preceq ,l_0,Inc)$ be a labeling structure. ${\cal L}$ is  {\em DCL (Detect new Cliques with Labels) } if 
for any integers $i$ and $n$ such that $1 \leq i < n$ and any subsets $I$ and $I'$ of $[i+2,n]$, 
if $I \subseteq I'$ and $lab_{\cal L}(I') \prec lab_{\cal L}(I \cup \{i+1\})$ then $I=I'$.
\end{Definition}

The labeling structures associated with MCS, LexBFS and MNS are clearly DCL, but  ${\cal L}_{LexDFS}$ is not since for 
any subsets $I$ and $I'$ of $[i+2,n]$, $lab_{\cal L}(I') \prec lab_{\cal L}(I \cup \{i+1\})$ necessarily holds.

\begin{Remark}
For each $X \in \{MCS, LexBFS, MNS\}$, ${\cal L}_X$ is a DCL labeling structure, but ${\cal L}_{LexDFS}$ is not.
\end{Remark}

\begin{algorithm}[H]
\SetKwInOut{Input}{input}
\SetKwInOut{Output}{output}
\textbf{Algorithm DCL-MLS-CliqueTree} 
\BlankLine
\Input{a connected chordal graph $H$ and a DCL labeling structure ${\cal L} = (L, \preceq,l_{0},Inc)$}
\Output{ a pmo $\alpha = (x_1, \ldots, x_n)$,
a clique tree $T$ and the set $Sep$ of minimal separators of $H$}
\BlankLine
{ % begin
\textbf{InitCT}\;
Initialize all labels as $l_{0}$; $prev$-$max$-$label \leftarrow l_0$\;
\For{ $i=n$  \textbf{downto} $1$}
{ % for
Choose a vertex $x$ in $V \setminus V'$ whose label is maximal, \\ 
and if possible strictly greater than $prev$-$max$-$label$\;
$x_i \leftarrow x$;  $S \leftarrow N_H(x) \cap V'$;
// $S = N_H^{\alpha+}(x)$ \\
\If {$prev$-$max$-$label \not \prec label(x)$ and $i <n$} 
{
\textbf{StartClique}\;
}
\textbf{IncreaseClique}\;
\ForEach {$y$ in $N_{H}(x) \setminus V'$}
{
$label(y) \leftarrow Inc(label(y),i)$\;
}
$V' \leftarrow V' \cup \{x\}$; $prev$-$max$-$label \leftarrow label(x)$\;
} % for
\textbf{DefineCT};
} % begin
\end{algorithm}

\begin{Theorem} \label{thDCL}
Algorithm DCL-MLS-CliqueTree is correct, and would be incorrect with any non-DCL input labeling structure.
Moreover, if the input labeling structure is ${\cal L}_X$
with $X \in \{MCS, LexBFS\}$ then the algorithm runs in linear time.
\end{Theorem}

\begin{Proof} 
We suppose that ${\cal L}$ is DCL.
It is sufficient to show that at each iteration $i$ in $[1,n-1]$
 $K_s = S \Leftrightarrow prev$-$max$-$label \prec label(x)$, \textit{i.e.} by Invariant~\ref{invMCT} e)
$N_H^{\alpha+}[x_{i+1}] = N_H^{\alpha ,i}(x_i) \Leftrightarrow prev$-$max$-$label \prec label(x)$.
The implication from left to right immediately follows from
 Lemma~\ref{lemMMLS} a).
Let us show the reverse implication.
We suppose that $ prev$-$max$-$label \prec label(x_i)$.
By Lemma~\ref{lemMMLS} b) $\{x_{i+1}\} \subseteq N_H^{\alpha ,i}(x_i) \subseteq N_H^{\alpha+}[x_{i+1}] $.
 Let $I = \alpha ^{-1}(N_H^{\alpha ,i+1}(x_i))$ and let $I' = \alpha ^{-1}(N_H^{\alpha+}(x_{i+1}))$.
$I \subseteq I' \subseteq [i+2,n]$ and 
$lab_{\cal L}(I') = prev$-$max$-$label \prec label(x_i) = lab_{\cal L}(I \cup \{i+1\})$, so $I=I'$ since ${\cal L}$ is DCL. 
It follows that $N_H^{\alpha+}[x_{i+1}] = N_H^{\alpha ,i}(x_i)$. \\
We suppose now that ${\cal L}$ is not DCL. 
Then there are some
integers $i$ and $n$ with $1 \leq i < n$ and some subsets $I$ and $I'$ of $[i+2,n]$ such that $I \subset I'$ 
and $lab_{\cal L}(I') \prec lab_{\cal L}(I \cup \{i+1\})$.
Let $H = (V,E)$ and $\alpha$ be defined by: $V = \{1, \ldots, n\}$, $\alpha = (1, \ldots, n)$, $\{ {i+2}, \ldots, n\}$ is a 
clique of $H$,  
$N_H^{\alpha+}({i+1}) = I'$ and $\forall j \in [1,i]$ $N_H^{\alpha+}({j}) = I \cup \{i+1\}$. 
$H$ is connected and chordal and by condition IC $\alpha$ can be computed by an execution of Algorithm DCL-MLS-CliqueTree. 
At iteration $i$ of such an execution, 
$K_s \neq S$ but $prev$-$max$-$label = lab_{\cal L}(I') \prec lab_{\cal L}(I \cup \{i+1\}) = label(x)$, so the execution 
increases current clique $K_s$ instead of starting a new one. \\
We suppose that the input labeling structure is ${\cal L}_X$ with $X \in \{MCS, LexBFS\}$.
As the order on labels is total, it is sufficient to choose a vertex $x$ with maximal label at each iteration.
As ${\cal L}_X$-MLS runs in linear time, it is sufficient to check that condition $prev$-$max$-$label \not \prec label(x)$ can 
be evaluated in $O(|N_H(x)|)$ time.
It is obviously the case if $X = MCS$. It is also the case if $X = LexBFS$ since $label(x)$ is of length at most  $|N_H(x)|$. 
%
%using as  data structure a list of vertex lists as described in \cite{RTL} where each vertex list contians 
%the vertices having the same label, condition $prev$-$max$-$label \not \prec label(x)$ holds when no new list has 
%been created during iteration $i+1$ from the list of vertices having $prev$-$max$-$label$ as label, which can be determined in $O(1)$ time.
%
\end{Proof} 

For $X = MCS$ (resp. $LexBFS$), as ${\cal L}_X$ is DCL with totally ordered labels, Algorithm DCL-${\cal L}_X$-MLS-CliqueTree can be 
simplified by choosing an arbitrary vertex with maximal label at each iteration, and we refind
the algorithms from \cite{BP} (resp. \cite{IPL}).
As ${\cal L}_{LexDFS}$ is not DCL, it follows from Theorem~\ref{thDCL} that
Algorithm DCL-MLS-CliqueTree would be incorrect with ${\cal L}_{LexDFS}$ as input labeling structure.
Note that this contradicts Theorem 4.1 from \cite{IPLLexDFS} stating that in an execution of LexDFS, $label(x_i) \preceq prev$-$max$-$label$ is 
a necessary and sufficient condition for $N_H^{\alpha+}[x_{i+1}]$ to be a maximal clique and $N_H^{\alpha+}(x_i)$ to be a minimal separator 
of the input graph, implying that DCL-MLS-CliqueTree is correct with ${\cal L}_{LexDFS}$ as input labeling structure. The simple graph $H$ 
from Figure~\ref{figchordal} is a counterexample as shown below.

\begin{figure}
\begin{center}
\begin{tikzpicture} 

%\begin{scope}
\coordinate (1) at (0,2) ;
\draw (1) node {$\bullet$}
               node  [above] {$a$ 1 (2,6)};

\coordinate (2) at (0,0) ;
\draw  (2) node {$\bullet$}
               node [below] {$b$ 2 (6)};
               
 \coordinate (3) at (5,2) ;
\draw  (3) node {$\bullet$}
               node  [above] {$c$ 3 (4,5)};
               
\coordinate (4) at (5,0) ;
\draw  (4) node {$\bullet$}
               node  [below] {$d$ 4 (5)};
                
\coordinate (5) at (4,1) ;
\draw  (5) node {$\bullet$}
               node  [below left] {$e$ 5 (6)};          
                
\coordinate (6) at (1,1) ;
\draw  (6) node {$\bullet$}
               node  [below right] {$f$ 6 ()};  
              
\draw (6) -- (1) -- (2) -- (6) -- (5) -- (3) -- (4) -- (5) ;

\draw  (0.4,1) node {$K$};
\draw  (4.6,1) node {$K'$};
%\end{scope}      

\end{tikzpicture}
\end{center} 

\caption{ LexDFS labels do not detect new maximal cliques.}
\label{figLexDFS}

\end{figure}
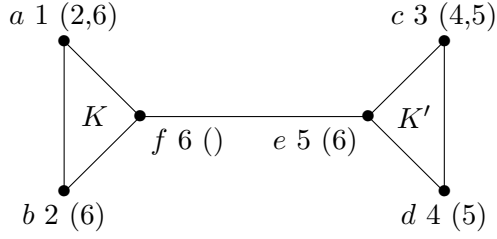

\begin{Counterexample}
An execution of Algorithm DCL-MLS-CliqueTree on the graph $H$ shown in Figure~\ref{figchordal} and the labeling structure ${\cal L}_{LexDFS}$ 
computing ordering $(a,b,c,d,e,f)$ is shown in Figure~\ref{figLexDFS}.
For each vertex $x$, the number $\alpha^{-1}(x)$ and the final label of $x$ are indicated.
At the beginning of iteration 4,  $label(a) = label(b) = pre$-$max$-$label = (6)$ and $label(c) = label(d) = (5)$, with $(6) \prec (5)$ according 
to labeling structure ${\cal L}_{LexDFS}$.  At iteration 4 vertex $d$ is chosen, and as $prev$-$max$-$label \prec label(d)$ the execution increases 
the current clique $\{e,f\}$ instead of starting new clique $K'$.
\end{Counterexample} 

%%%%%%%%%%%%%%%%%%%%%%%%%%%%%%%

%\begin{figure}
%\centerline{\includegraphics[width=8cm]{fig2}}
%\caption{ LexDFS labels do not detect new maximal cliques.}
%\label{figLexDFS}
%\end{figure} 

%%%%%%%%%%%%%%%%%%%%%%%%%%%%%%%

\section{Clique tree of the complement graph}

Algorithm MLS can be used to compute a peo of the complement graph \cite{GLS}. We will show that it can be used to compute a pmo, a clique tree 
and the minimal separators of the complement graph, provided that this complement graph is connected and chordal.

\begin{Definition}
Let ${\cal L}=(L,\preceq ,l_0,Inc)$ be a labeling structure. ${\cal L}$ is  {\em complement-reversing} if for any integers $i$ and $n$ 
with $1 \leq i \leq n$ and any subsets $I$ and $I'$ of $[i,n]$, if $lab_{\cal L}(I) \preceq lab_{\cal L}(I')$ then $lab_{\cal L}([i,n] \setminus I') \preceq lab_{\cal L}([i,n] \setminus I)$.
\end{Definition}

\begin{Remark} \cite{GLS} \label{remRev}
For each $X \in \{MCS, LexBFS, LexDFS, MNS\}$, ${\cal L}_X$ is complement-reversing.
\end{Remark}

\begin{Definition}
For any labeling structure  ${\cal L}$, $Rev({\cal L})$ denotes the labeling structure obtained from ${\cal L}$ by replacing the order on the 
labels by its dual order.
\end{Definition}

\begin{Theorem} \cite{GLS} \label{thGLS}
Let ${\cal L}$ be a complement-reversing labeling structure. Then an ordering of a graph $G$ is a ${\cal L}$-MLS ordering of $\overline{G}$ 
if and only if it is a $Rev({\cal L})$-MLS ordering of $G$.
\end{Theorem}

Thus if ${\cal L}$ is complement-reversing then replacing "maximal" by "minimal" in algorithm MLS makes an execution of the obtained algorithm 
with $G$ and ${\cal L}$ as input compute a $Rev({\cal L})$-MLS ordering of $G$ and therefore a ${\cal L}$-MLS ordering of $\overline{G}$, 
which is a peo of $\overline{G}$ if it is chordal.
However, it is not correct to replace in Algorithm Moplex-MLS or MLS-CliqueTree condition ``if possible strictly greater than $prev$-$max$-$label$'' 
by ``if possible strictly smaller than $prev$-$max$-$label$'' since in case $N_H^{\alpha+}[x_{i+1}] = N_H^{\alpha+}(x_i)$. With $H = \overline{G}$, 
we have  $N_G^{\alpha+}(x_{i+1}) = N_G^{\alpha+}(x_i)$ \textit{i.e.} $prev$-$min$-$label = label(x)$ at iteration $i$ (as $x$ 
is adjacent to $x_{i+1}$ in $H$, it is not adjacent to $x_{i+1}$ in $G$ and therefore its label is not increased during iteration $i+1$). 
We will show that $prev$-$min$-$label \neq label(x)$ is a necessary and sufficient condition for starting a new clique in $H$, whether the 
labeling structure is DCL or not.

\begin{algorithm}[H]
\SetKwInOut{Input}{input}
\SetKwInOut{Output}{output}
\textbf{Algorithm Complement-MLS-CliqueTree} 
\BlankLine
\Input{a graph $G$ whose complement is a connected chordal graph and a complement-reversing labeling structure ${\cal L} = (L, \preceq,l_{0},Inc)$}
\Output{ a pmo $\alpha = (x_1, \ldots, x_n)$,
a clique tree $T$ and the set $Sep$ of minimal separators of $\overline{G}$}
\BlankLine
{ % begin
$H \leftarrow \overline{G}$; \textbf{InitCT}\; 
Initialize all labels as $l_{0}$; $prev$-$min$-$label \leftarrow l_0$\;
\For{ $i=n$  \textbf{downto} $1$}
{ % for
Choose a vertex $x$ in $V \setminus V'$ whose label is minimal, and if possible equal to $prev$-$min$-$label$\;
$x_i \leftarrow x$;  $S \leftarrow N_H(x) \cap V'$;
// $S = N_H^{\alpha+}(x)$ \\
\If {$prev$-$min$-$label \neq label(x)$ and $i<n$} 
{
\textbf{StartClique}; $prev$-$min$-$label \leftarrow label(x)$\;

}
\textbf{IncreaseClique}\;
\ForEach {$y$ in $N_{G}(x) \setminus V'$}
{
$label(y) \leftarrow Inc(label(y),i)$\;
}
$V' \leftarrow V' \cup \{x\}$\;
} % for
\textbf{DefineCT};
} % begin
\end{algorithm}

Note that as by condition IC labels can only increase, $prev$-$min$-$label$ is necessarily minimal.

\begin{Theorem} \label{thCompCT}
Algorithm Complement-MLS-CliqueTree is correct.
\end{Theorem}

 An ordering computed by Algorithm Complement-MLS-CliqueTree is a $Rev({\cal L})$-MLS ordering of $G$ and therefore by Theorem~\ref{thGLS} 
a ${\cal L}$-MLS ordering of $H$, which is a peo of $H$.
To prove  Theorem~\ref{thCompCT} we will use the following Lemma.

\begin{Lemma} \label{lemCompCT}
In an execution of Complement-MLS-CliqueTree, 
for each $i$ in $[1,n-1]$ and each $y$ in $V$ such that $\alpha^{-1}(y) \leq i$, 
at the beginning of iteration $i$, the following propositions are equivalent:

\begin{itemize}
\item[1)] $N_H^{\alpha+}[x_{i+1}] \subseteq N_H^{\alpha ,i}(y)$,
\item[2)] $prev$-$min$-$label = label(y)$,
\item[3)] $N_H^{\alpha+}[x_{i+1}] = N_H^{\alpha ,i}(y)$.
\end{itemize}
\end{Lemma}

\begin{Proof} 
1) $\Rightarrow$ 2): 
We suppose that $N_H^{\alpha+}[x_{i+1}] \subseteq N_H^{\alpha ,i}(y)$. Then $N_G^{\alpha ,i}(y) \subseteq N_G^{\alpha ,i}(x_{i+1})$, and 
therefore by Lemma~\ref{lemIC}
 $label(y) \preceq label(x_{i+1}) = prev$-$min$-$label$. Moreover $label(y) \not \prec prev$-$min$-$label$ since $label(y)$ 
and $prev$-$min$-$label$ are the labels of $y$ and $ x_{i+1}$ respectively at the beginning of iteration $i+1$. 
Hence $prev$-$min$-$label = label(y)$. \\
2) $\Rightarrow$ 3): 
We suppose that $prev$-$min$-$label = label(y)$. Then the label of $y$ is not increased during iteration $i+1$ (otherwise it would have been 
strictly smaller than the label of $x_{i+1}$ at the beginning of iteration $i+1$). It follows that $y$ is not adjacent to $x_{i+1}$ in $G$, 
and therefore is adjacent to $x_{i+1}$ in $H$. As $\alpha$ is a peo of $H$, $N_H^{\alpha ,i}(y) \subseteq N_H^{\alpha+}[x_{i+1}]$.
Moreover $N_H^{\alpha ,i}(y) \not \subset N_H^{\alpha+}[x_{i+1}]$ since otherwise $N_G^{\alpha ,i}(x_{i+1}) \subset N_G^{\alpha ,i}(y)$ and 
therefore by Lemma~\ref{lemIC}
 $prev$-$min$-$label \prec label(y)$.
Hence $N_H^{\alpha+}[x_{i+1}] = N_H^{\alpha ,i}(y)$. \\
3) $\Rightarrow$ 1) is evident.
\end{Proof} 
	
\begin{Proof} (of Theorem~\ref{thCompCT})
Let $\alpha = (x_1, \ldots, x_n)$ be the ordering computed by an execution of Complement-MLS-CliqueTree on input $G$ and ${\cal L}$, 
and let $H = \overline{G}$.
Let us show that it is a pmo of $H$, \textit{i.e.} a MCComp peo of $H$
by Characterization~\ref{carMoplex}.
As
$\alpha$ is a MLS ordering of $H$, it is a peo of $H$.
Let $i \in [1,n-1]$. We suppose that $N_H^{\alpha+}[x_{i+1}]$ is not a maximal clique of $H$. Let us show that it is equal 
to $N_H^{\alpha+}(x_{i})$. 
As $N_H^{\alpha+}[x_{i+1}]$  is not a maximal clique of $H$, there is a vertex $y$ such that $\alpha^{-1}(y) \leq i$ 
and $N_H^{\alpha+}[x_{i+1}] \subseteq N_H^{\alpha ,i}(y)$, and therefore by Lemma~\ref{lemCompCT} 
$prev$-$min$-$label = label(y)$
 at the beginning of iteration $i$ in this execution. 
It follows by the condition on the choice of $x$
 that $prev$-$min$-$label = label(x_{i})$, and therefore by Lemma~\ref{lemCompCT}
that $N_H^{\alpha+}[x_{i+1}] = N_H^{\alpha+}(x_i)$. \\
It remains to show that condition $prev$-$min$-$label \neq label(x)$ correctly detects new cliques. It is evident at iteration $n$. 
For each iteration $i$ with $i<n$ it immediately follows from Lemma~\ref{lemCompCT}, as the condition to start a new clique is $K_s \neq S$, 
\textit{i.e.} $N_H^{\alpha+}[x_{i+1}] \neq N_H^{\alpha ,i}(x)$ by Invariant~\ref{invMCT} e).
\end{Proof} 

\begin{figure}
\begin{center}
\begin{tikzpicture} 

%\begin{scope}
\coordinate (1) at (0,2) ;
\draw (1) node {$\bullet$}
               node  [above] {$a$ 1 (3,4,5)};

\coordinate (2) at (0,0) ;
\draw  (2) node {$\bullet$}
               node [below] {$b$ 2 (3,4,5)};
               
 \coordinate (3) at (5,2) ;
\draw  (3) node {$\bullet$}
               node  [above] {$c$ 3 (6)};
               
\coordinate (4) at (5,0) ;
\draw  (4) node {$\bullet$}
               node  [below] {$d$ 4 (6)};
                
\coordinate (5) at (4,1) ;
\draw  (5) node {$\bullet$}
               node  [right] {$e$ 5 ()};          
                
\coordinate (6) at (1,1) ;
\draw  (6) node {$\bullet$}
               node  [left] {$f$ 6 ()};  
              
\draw (1) -- (5) -- (2) -- (4) -- (6) -- (3) -- (1) -- (4) ;
\draw (2) -- (3)  ;

%\end{scope}      
             
\end{tikzpicture}
\end{center} 

\caption{ LexDFS labels detect new cliques on the complement.}
\label{figComplementLexDFS}

\end{figure}
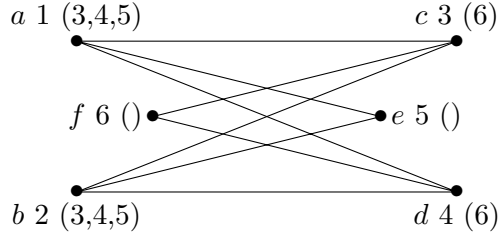

\begin{Example}
Let $G$ be the complement graph of the graph $H$ shown in Figure~\ref{figchordal}.
An execution of Algorithm Complement-MLS-CliqueTree on $G$ and labeling structure ${\cal L}_{LexDFS}$ computing ordering $(a,b,c,d,e,f)$ 
is shown in Figure~\ref{figComplementLexDFS}.
For each vertex $x$, the number $\alpha^{-1}(x)$ and the final label of $x$ are indicated.
At the beginning of iteration 4,  $label(a) = label(b) = (5)$ and $label(c) = label(d) = (6)$,with $(6) \prec (5)$ according to labeling 
structure ${\cal L}_{LexDFS}$, and $pre$-$min$-$label = ()$;  vertex $d$ is chosen and new clique $K'$ is started at iteration 4 as desired. 
Clique $K$ is correctly started at iteration 2 since $pre$-$min$-$label = (6)$ and $label(x) = label(b) = (3,4,5)$.
\end{Example} 

%%%%%%%%%%%%%%%%%%%%%%%%%%%%%%%

%\begin{figure}
%\centerline{\includegraphics[width=8.3cm]{fig3}}
%\caption{ LexDFS labels detect new cliques on the complement.}
%\label{figComplementLexDFS}
%\end{figure} 

%%%%%%%%%%%%%%%%%%%%%%%%%%%%%%%

Algorithm Complement-MLS-CliqueTree  does not run in $O(n+m)$ time, since computing $S$ (in order to compute the maximal cliques and 
minimal separators of $\overline{G}$) and $k$ in \textbf{StartClique} (in order to compute the edges of the clique tree) globally 
takes $O(n+\overline{m})$ time where $\overline{m}$ is the number of edges of $\overline{G}$. However, it is possible to compute a 
pmo $\alpha$ and the generators of the maximal cliques and minimal separators w.r.t. $\alpha$ of $\overline{G}$ in $O(n+m)$ time.

\begin{algorithm}[H]
\SetKwInOut{Input}{input}
\SetKwInOut{Output}{output}
\textbf{Algorithm Complement-MLS-Generators} 
\BlankLine
\Input{a graph $G$ whose complement is a connected chordal graph and a complement-reversing labeling 
structure ${\cal L} = (L, \preceq,l_{0},Inc)$}
\Output{ a pmo $\alpha = (x_1, \ldots, x_n)$ and the sets $GenCli$ and $GenSep$ of generators of the maximal 
cliques and minimal separators w.r.t. $\alpha$ of $\overline{G}$ respectively}
\BlankLine
{ % begin
$V' \leftarrow \emptyset$; $GenCli \leftarrow \emptyset$;$GenSep \leftarrow \emptyset$\; 
Initialize all labels as $l_{0}$; $prev$-$min$-$label \leftarrow l_0$\;
\For{ $i=n$  \textbf{downto} $1$}
{ % for
Choose a vertex $x$ in $V \setminus V'$ whose label is minimal, and if possible equal to $prev$-$min$-$label$\;
$x_i \leftarrow x$\;
\If {$prev$-$min$-$label \neq label(x)$ and $i<n$} 
{
$GenCli \leftarrow GenCli + \{x_{i+1}\}$;
$GenSep \leftarrow GenSep + \{x_{i}\}$\;
$prev$-$min$-$label \leftarrow label(x)$\;
}
\ForEach {$y$ in $N_{G}(x) \setminus V'$}
{
$label(y) \leftarrow Inc(label(y),i)$\;
}
$V' \leftarrow V' \cup \{x\}$\;
} % for
$GenCli \leftarrow GenCli + \{x_{1}\}$;
} % begin
\end{algorithm}

\begin{Theorem} \label{thCompGen}
Algorithm Complement-MLS-Generators is correct, and if the input labeling structure is ${\cal L}_X$
with $X \in \{MCS, LexBFS\}$ then it runs in linear time. 
\end{Theorem}

\begin{Proof} 
Correctness follows from the correctness of Complement-MLS-CliqueTree. \\
We suppose that the input labeling  structure is ${\cal L}_X$ with $X \in \{MCS, LexBFS\}$.
As the order on the labels is total, it is sufficient to choose a vertex $x$ with minimal label at each iteration.
Linear time complexity of  ${\cal L}_X$-MLS also holds for $Rev({\cal L}_X)$-MLS.
Hence it is sufficient to check that condition $prev$-$min$-$label \neq label(x)$ can be evaluated in $O(|N_G(x)|)$ time.
It is obviously the case if $X = MCS$. It is also the case if $X = LexBFS$ since $label(x)$ is of length at most  $|N_G(x)|$. 
%
%$O(1)$ time. It is obviously the case if $X = MCS$. It is also the case if $X = LexBFS$: using as  data structure a list of 
%vertex lists as described in \cite{RTL} where each vertex list contains the vertices having the same label, 
%condition $prev$-$min$-$label \neq label(x)$ holds when the list of vertices having $prev$-$min$-$label$ as 
%label has become empty during previous iteration, which can be detetermined in $O(1)$ time.
%
\end{Proof} 

%\newpage

\section{Extended results}

\subsection{Clique tree of a minimal triangulation}

Algorithm MLSM computes a meo and the associated minimal triangulation of the input graph $G$. 
It computes a mmo of $G$ if the order on labels is total \cite{Extremities}.
It can be modified into Algorithms Moplex-MLSM computing a mmo of $G$ whether the order on labels is total or not, 
which can be extended to Algorithms MLSM-CliqueTree
and DCL-MLSM-CliqueTree computing  a mmo, the associated minimal triangulation $H$ of $G$, a clique tree and the minimal 
separators of $H$. Below are algorithms Moplex-MLSM and
DCL-MLSM-CliqueTree.

\begin{algorithm}[H]
\SetKwInOut{Input}{input}
\SetKwInOut{Output}{output}
\textbf{Algorithm Moplex-MLSM} 
\BlankLine
\Input{a connected graph $G$ and a labeling structure ${\cal L} = (L, \preceq,l_{0},Inc)$}
\Output{a mmo $\alpha = (x_1, \ldots, x_n)$ of $G$ and the associated minimal triangulation $H = G^+_{\alpha}$}
\BlankLine
{ % begin
$H \leftarrow G$; $V' \leftarrow \emptyset$; Initialize all labels as $l_{0}$; $prev$-$max$-$label \leftarrow l_0$\;
\For{ $i=n$  \textbf{downto} $1$}
{ % for1
Choose a vertex $x$ in $V \setminus V'$ whose label is maximal, \\
and if possible strictly greater than $prev$-$max$-$label$\;
$x_i \leftarrow x$\; 
$Y \leftarrow \emptyset$\;
\ForEach {$y$ in $V \setminus V'$}
{ % for2
\If {there is a path $\mu$ of length $\geq 1$    in $G(V \setminus V')$ between $x$ and $y$ such that for each internal 
vertex $z$ of $\mu$ $label(z) \prec label(y)$} 
{ % if
$Y \leftarrow Y + \{y\}$; $E(H) \leftarrow E(H) \cup \{xy\}$\;
} % if
} % for2
\ForEach {$y$ in $Y$}
{
$label(y) \leftarrow Inc(label(y),i)$\;
}
$V' \leftarrow V' \cup \{x\}$; $prev$-$max$-$label \leftarrow label(x)$\;
} % for1
} % begin
\end{algorithm}

\begin{Theorem} \label{thMMLSM}
Algorithm Moplex-MLSM computes a mmo and the associated minimal triangulation of the input graph.
\end{Theorem}

To prove Theorem~\ref{thMMLSM} we will use the following Lemmas.

\begin{Lemma} \label{lemMMLSM1} \cite{BBDirac}
A moplex of a minimal triangulation of $G$ is a moplex of $G$.
\end{Lemma}

\begin{Lemma} \label{lemMMLSM2} 
If $\alpha$ is a meo of $G$ and a pmo of $G^+_{\alpha}$ then it is a mmo of $G$.
\end{Lemma}

\begin{Proof}
Let $H = G^+_{\alpha}$.
We prove this by induction on the size $k$ of the perfect moplex partition $(X_1, \ldots, X_k)$ associated with $\alpha$ in $H$.
If $k = 1$ then $H$ is a clique, so $G$ is a clique too since $H$ is a minimal triangulation of $G$, and we are done.
We assume that the property holds for a perfect moplex partition of size $k \geq 1$. Let $(X_1, \ldots, X_{k+1})$ be the perfect 
moplex partition associated with $\alpha$ in $H$.
As $X_1$ is a moplex of $H$, by Lemma~\ref{lemMMLSM1} it is a moplex of $G$. Let $G_1$ be the graph obtained from $G$ by saturating $N_G(X_1)$ and removing $X_1$, and let $\alpha_1$ be the restriction of $\alpha$ to $V \setminus X_1$.

As $\alpha$ is a meo of $G$, $\alpha_1$ is a meo of $G_1$, and
as $(G_1)^+_{\alpha_1} = H(V \setminus X_1)$, $\alpha_1$ is a pmo of $(G_1)^+_{\alpha_1}$.
Hence by induction hypothesis $\alpha_1$ is a mmo of $G_1$, and therefore $\alpha$ is a mmo of $G$.
\end{Proof}

Note that the fact that $\alpha$ is a pmo of $G^+_{\alpha}$ does not imply that it is a mmo of $G$. For instance if $G$ is a non-clique graph 
with a universal vertex then any ordering $\alpha$ of $G$ such that $\alpha(1)$ is universal is a pmo of $G^+_{\alpha}$ (since $G^+_{\alpha}$ 
is a clique) but not a mmo of $G$ (since it is not a meo of $G$).

\begin{Proof} (of Theorem~\ref{thMMLSM} )
Let $\alpha$ be the ordering and $H$ be the graph computed by an execution of Moplex-MLSM on input graph $G$. As 
this execution is also an execution of MLSM, $\alpha$ is
a meo of $G$ and $H = G^+_{\alpha}$. As moreover at each iteration the labels are increased exactly in the same way as in an execution of 
Moplex-MLS on $H$, $\alpha$ is a pmo of $H$ and therefore a mmo of $G$ by Lemma~\ref{lemMMLSM2}.
\end{Proof}

\begin{algorithm}[H]
\SetKwInOut{Input}{input}
\SetKwInOut{Output}{output}
\textbf{Algorithm DCL-MLSM-CliqueTree} 
\BlankLine
\Input{a connected graph $G$ and a DCL labeling structure ${\cal L} = (L, \preceq,l_{0},Inc)$}
\Output{ a mmo $\alpha = (x_1, \ldots, x_n)$ of $G$, the associated minimal triangulation $H = G^+_{\alpha}$,
a clique tree $T$ and the set $Sep$ of minimal separators of $H$}
\BlankLine
{ % begin
$H \leftarrow G$; \textbf{InitCT}\;
Initialize all labels as $l_{0}$; $prev$-$max$-$label \leftarrow l_0$\;
\For{ $i=n$  \textbf{downto} $1$}
{ % for
Choose a vertex $x$ in $V \setminus V'$ whose label is maximal, \\ 
and if possible strictly greater than $prev$-$max$-$label$\;
$x_i \leftarrow x$;  $S \leftarrow N_H(x) \cap V'$;
// $S = N_H^{\alpha+}(x)$ \\
\If {$prev$-$max$-$label \not \prec label(x)$ and $i<n$} 
{
\textbf{StartClique}\;
}
\textbf{IncreaseClique}\;

$Y \leftarrow \emptyset$\;
\ForEach {$y$ in $V \setminus V'$}
{ % for2
\If {there is a path $\mu$ of length $\geq 1$    in $G(V \setminus V')$ between $x$ and $y$ such that for each internal vertex $z$ 
of $\mu$ $label(z) \prec label(y)$} 
{ % if

$Y \leftarrow Y + \{y\}$; $E(H) \leftarrow E(H) \cup \{xy\}$\;
} % if
} % for2
\ForEach {$y$ in $Y$}
{
$label(y) \leftarrow Inc(label(y),i)$\;
}
$V' \leftarrow V' \cup \{x\}$; $prev$-$max$-$label \leftarrow label(x)$\;
} % for
\textbf{DefineCT};
} % begin
\end{algorithm}

Correctness of Algorithm DCL-MLSM-CliqueTree  immediately follows from the correctness of Algorithms Moplex-MLSM and DCL-MLS-CliqueTree.
 
\subsection{Atom tree and clique minimal separators}

An atom tree of a connected graph $G$ can be computed from a clique tree of a minimal triangulation of $G$ as described in the following Theorem.

\begin{Theorem} 
 \label{thMerge} \cite{AtomTree}
Let $G$ be a connected graph, let $H$ be a minimal triangulation of $G$, let $T = (V_T,E_T)$ 
be a clique tree of $H$, and let $T'$ be the forest obtained from $T$ by removing all edges $KK'$ such that $K \cap K'$ 
is a clique in $G$,
let $T''$ be the tree obtained from $T$ by merging the nodes of each tree of $T'$ into one node; then 
$T''$ is an atom tree of $G$, and for each edge $KK'$ of $T$ such that $K \cap K'$ 
is a clique in $G$, $K \cap K' = A \cap A'$, where $A$ and $A'$ are the atoms of $G$ containing $K$ and $K'$ respectively.
\end{Theorem}

Thus Algorithm DCL-MLSM-CliqueTree  can be modified into Algorithm DCL-AtomTree  computing an atom tree and the clique minimal separators 
of the input graph $G$: in case a new clique is started, a new atom is started only if $S$ is a clique in $G$, otherwise the atom 
containing $K_p$ is increased. Note that the atoms are not built one after the other since an atom different from $A_s$ may be increased 
if a new clique is started and $S$ is not a clique in $G$. Variable $q$ contains the index of the current atom. Algorithm DCL-AtomTree  
generalizes algorithm MCSM-Atom-Tree  from \cite{AtomTree}, while correcting an error in this algorithm (a confusion between 
variables $q$ and $s$)..

\begin{algorithm}[H]
\textbf{InitAT} \\
%\BlankLine 
{
$V' \leftarrow \emptyset$; $s \leftarrow 1$; $q \leftarrow 1$; $A_1 \leftarrow \emptyset$; $E \leftarrow \emptyset$; $CliqueSep \leftarrow \emptyset$; 
}
\end{algorithm}

\begin{algorithm}[H]
\textbf{StartAtom} \\
%\BlankLine 
{
$s \leftarrow s+ 1; A_s \leftarrow S$\; 
$E \leftarrow E \cup \{(p,s)\} $;
$CliqueSep\leftarrow CliqueSep \cup \{S\}$;
}
\end{algorithm}

\begin{algorithm}[H]
\SetKwInOut{Input}{input}
\textbf{IncreaseAtom} \\
%\BlankLine
\Input{an integer $q$}
{
$A_q \leftarrow A_q \cup \{x\}$; $atom(x) \leftarrow q$;  // increase atom $A_q$
}
\end{algorithm}

\begin{algorithm}[H]
\textbf{DefineAT} \\
%\BlankLine 
{
$T \leftarrow (\{A_1, \dots, A_s\},  \{A_p A_q,~(p,q) \in E\})$;
}
\end{algorithm}

\begin{algorithm}[H]
\SetKwInOut{Input}{input}
\SetKwInOut{Output}{output}
\textbf{Algorithm DCL-AtomTree} 
\BlankLine
\Input{a connected graph $G$ and a DCL labeling structure ${\cal L} = (L, \preceq,l_{0},Inc)$}
\Output{ an atom tree $T$ and the set $CliqueSep$ of clique minimal separators of $G$}
\BlankLine
{ % begin
$H \leftarrow G$; \textbf{InitAT}\;
Initialize all labels as $l_{0}$; $prev$-$max$-$label \leftarrow l_0$\;
\For{ $i=n$  \textbf{downto} $1$}
{ % for
Choose a vertex $x$ in $V \setminus V'$ whose label is maximal, \\ 
and if possible strictly greater than $prev$-$max$-$label$\;
$x_i \leftarrow x$;  $S \leftarrow N_H(x) \cap V'$;
// $S = N_H^{\alpha+}(x)$ \\
\If {$prev$-$max$-$label \not \prec label(x)$ and $i<n$} 
{
$k \leftarrow min\{j, \alpha(j) \in S\}$;
$p \leftarrow clique(\alpha(k)) $\; 
\If {$S$ is a clique in $G$}
{
\textbf{StartAtom}; $q \leftarrow s$\;
}
\Else
{
$q \leftarrow p$\;
}
}
\textbf{IncreaseAtom}$(q)$\;
$Y \leftarrow \emptyset$\;
\ForEach {$y$ in $V \setminus V'$}
{ % for2
\If {there is a path $\mu$ of length $\geq 1$ in $G(V \setminus V')$ between $x$ and $y$ such that for each internal 
vertex $z$ of $\mu$ $label(z) \prec label(y)$} 
{ % if
$Y \leftarrow Y + \{y\}$; $E(H) \leftarrow E(H) \cup \{xy\}$\;
} % if
} % for2
\ForEach {$y$ in $Y$}
{
$label(y) \leftarrow Inc(label(y),i)$\;
}
$V' \leftarrow V' \cup \{x\}$; $prev$-$max$-$label \leftarrow label(x)$\;
} % for
\textbf{DefineAT};
} % begin
\end{algorithm}

Correctness of Algorithm DCL-AtomTree follows from Theorem~\ref{thMerge}, Characterization~\ref{carAtomTree} and from the correctness 
of Algorithm DCL-MLSM-CliqueTree.

\subsection{MLS on a non-chordal graph}

A MLS ordering of a non-chordal graph $G$ is not necessarily a meo of $G$. Berry and Bordat \cite{BBDirac} showed that 
LexBFS ends on a moplex, \textit{i.e.}
if $\alpha$ is a LexBFS ordering of a non-clique graph $G$ then there is a moplex $X_1$ of $G$ such that 
$X_1 = \{\alpha(1), \ldots, \alpha(|X_1|)\}$. As the restriction of $\alpha$ to $V \setminus X_1$ is a LexBFS ordering 
of $G(V \setminus X_1)$ it follows that $\alpha$ is compatible with a simple moplex partition of $G$.
However, if $G$ is not chordal then $\alpha$ is not necessarily a pmo of $G$, and it is not necessarily a mmo of $G$, and 
not even a meo of $G$.

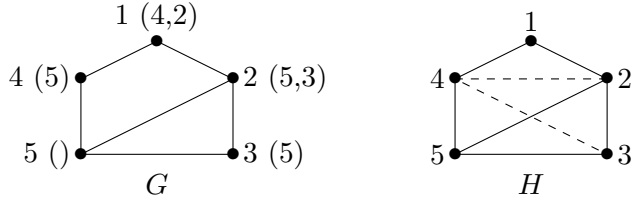
\begin{figure}
\begin{center}
\begin{tikzpicture} 

\begin{scope}
\coordinate (1) at (1,1.5) ;
\draw (1) node {$\bullet$}
               node  [above] {1 (4,2)};

\coordinate (2) at (2,1) ;
\draw  (2) node {$\bullet$}
               node [right] {2 (5,3)};
               
 \coordinate (3) at (2,0) ;
\draw  (3) node {$\bullet$}
               node  [right] {3 (5)};
               
\coordinate (4) at (0,1) ;
\draw  (4) node {$\bullet$}
               node  [left] {4 (5)};
                
\coordinate (5) at (0,0) ;
\draw  (5) node {$\bullet$}
               node  [left] {5 ()};          
              
\draw (2) -- (5) -- (3) -- (2) -- (1) -- (4) -- (5) ;

\draw (1,-0.4)  node {$G$};
\end{scope}      
             
\begin{scope}[xshift = 140]
\coordinate (1) at (1,1.5) ;
\draw (1) node {$\bullet$}
               node  [above] {1};

\coordinate (2) at (2,1) ;
\draw  (2) node {$\bullet$}
               node [right] {2};
               
 \coordinate (3) at (2,0) ;
\draw  (3) node {$\bullet$}
               node  [right] {3};
               
\coordinate (4) at (0,1) ;
\draw  (4) node {$\bullet$}
               node  [left] {4};
                
\coordinate (5) at (0,0) ;
\draw  (5) node {$\bullet$}
               node  [left] {5};          
              
\draw (2) -- (5) -- (3) -- (2) -- (1) -- (4) -- (5) ;

\draw [dash] (2) -- (4) -- (3) ; 

\draw (1,-0.4)  node {$H$};
\end{scope}

\end{tikzpicture}
\end{center} 

\caption{LexBFS on a non-chordal graph computes an ordering that is compatible with a simple moplex partition, but not a meo.}
\label{figLexBFS}

\end{figure}

\begin{Example}
Let $G$ be the graph shown in Figure~\ref{figLexBFS} and let $\alpha = (1,2,3,4,5)$.
$\alpha$ is a LexBFS ordering of $G$ (the final labels are indicated). $\alpha$ is compatible with the simple moplex partition
$(\{1\}, \{2,3\}, \{4,5\})$.
However, $\alpha$ is not a meo of $G$ since the graph $H = G^+_{\alpha}$ (shown in the figure with dashed fill edges) is not a 
minimal triangulation of $G$.
\end{Example} 

%%%%%%%%%%%%%%%%%%%%%%%%%%%%%%%

%\begin{figure}
%\centerline{\includegraphics[width=9.3cm]{fig4}}
%\caption{LexBFS on a non-chordal graph computes an ordering that is compatible with a simple moplex partition, but not a meo.}
%\label{figLexBFS}
%\end{figure} 

%%%%%%%%%%%%%%%%%%%%%%%%%%%%%%%

Berry and Bordat \cite{BBDirac} also showed that if $\alpha$ is a LexBFS ordering of a graph $G$ then the minimal separators included 
in $N(\alpha (1))$ are totally ordered by inclusion, and that $\alpha$ consecutively numbers the vertices of each connected component 
of $G(V \setminus N[\alpha (1)])$ and its neighborhood. More accurately there is an order $(C_1, \ldots, C_p)$ on the connected component 
of $G(V \setminus N[\alpha (1)])$ such that for each $i$ in $[1, p-1]$; $N(C_i) \subseteq N(C_{i+1})$ (the minimal separators 
included in $N(\alpha (1))$ are the sets $N(C_i)$) and $\alpha$ is compatible with the ordered 
partition $(X_1, N[C_p] \setminus N(C_{p-1}), \ldots, N[C_2] \setminus N(C_{1}), N[C_1])$, where $X_1$ is the moplex 
containing $\alpha (1)$.
Xu et al \cite{IPLLexDFS} showed that LexDFS also ends on a moplex of the input graph $G$ (and therefore a LexDFS ordering 
is compatible with a simple moplex partition of $G$) but they left open the question whether  LexDFS orderings also have the 
properties on minimal separators and  connected components.
The following counterexample shows that these properties of LexBFS do not extend to LexDFS.

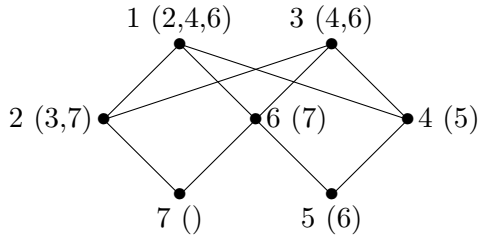
\begin{figure}
\begin{center}
\begin{tikzpicture} 

%\begin{scope}
\coordinate (1) at (1,2) ;
\draw (1) node {$\bullet$}
               node  [above] {1 (2,4,6)};

\coordinate (2) at (0,1) ;
\draw  (2) node {$\bullet$}
               node [left] {2 (3,7)};
               
 \coordinate (3) at (3,2) ;
\draw  (3) node {$\bullet$}
               node  [above] {3 (4,6)};
               
\coordinate (4) at (4,1) ;
\draw  (4) node {$\bullet$}
               node  [right] {4 (5)};
                
\coordinate (5) at (3,0) ;
\draw  (5) node {$\bullet$}
               node  [below] {5 (6)};          
              
\coordinate (6) at (2,1) ;
\draw  (6) node {$\bullet$}
               node  [right] {6 (7)};
                
\coordinate (7) at (1,0) ;
\draw  (7) node {$\bullet$}
               node  [below] {7 ()}; 
  
\draw (4) -- (1) -- (2) -- (7) -- (6) -- (1) ;
\draw (2) -- (3) -- (4) -- (5) -- (6) -- (3) ;

%\end{scope}      

\end{tikzpicture}
\end{center} 

\caption{With LexDFS, the minimal separators included in $N(\alpha (1))$ are not totally ordered by inclusion.}
\label{figLexDFScont1}

\end{figure}

\begin{figure}
\begin{center}
\begin{tikzpicture} 

%\begin{scope}
\coordinate (1) at (0,2) ;
\draw (1) node {$\bullet$}
               node  [above] {1 (5)};

\coordinate (2) at (0,-1) ;
\draw  (2) node {$\bullet$}
               node [below] {2 (3,6)};
               
 \coordinate (3) at (0.5,0) ;
\draw  (3) node {$\bullet$}
               node  [right] {3 (5)};
               
\coordinate (4) at (2,0) ;
\draw  (4) node {$\bullet$}
               node  [right] {4 (5)};
                
\coordinate (5) at (0,1) ;
\draw  (5) node {$\bullet$}
               node  [right] {5 (6)};          
              
\coordinate (6) at (-0.5,0) ;
\draw  (6) node {$\bullet$}
               node  [left] {6 ()};
                
\draw (1) -- (5) -- (6) -- (2) -- (3) -- (5) -- (4) ;

%\end{scope}      

\end{tikzpicture}
\end{center} 

\caption{LexDFS does not number the connected components of $G(V \setminus N[\alpha(1)])$ one after the other.}
\label{figLexDFScont2}

\end{figure}
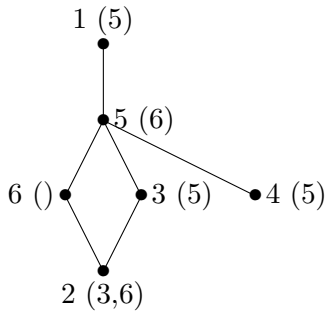

\begin{Counterexample}
Let $G$ be the graph shown in Figure~\ref{figLexDFScont1} and let $\alpha = (1,2,\ldots,,7)$.
$\alpha$ is a LexDFS ordering of $G$ (the final labels are indicated). The minimal separators included in $N[1]$ are $\{2,6\}$ and $\{4,6\}$, 
and are incomparable for inclusion. If the property on connected components was true, $C_1$ would be the component numbered last by $\alpha$, 
\textit{i.e.}  $\{7\}$, but the vertices of $N[\{7\}]$ ($7$, $6$ and $2$) are not numbered consecutively by $\alpha$.
If $G$ is the graph shown in Figure~\ref{figLexDFScont2} with $\alpha = (1,2,\ldots,,6)$, even the restriction of $\alpha$ to $V \setminus N[1]$ 
is not compatible with a sequence of the connected components of $G(V \setminus N[1])$.
\end{Counterexample} 

%%%%%%%%%%%%%%%%%%%%%%%%%%%%%%%

%\begin{figure}
%\centerline{\includegraphics[width=5cm]{fig5}}
%\caption{With LexDFS, the minimal separators included in $N(\alpha (1))$ are not totally ordered by inclusion.}
%\label{figLexDFScont1}
%\end{figure} 

%%%%%%%%%%%%%%%%%%%%%%%%%%%%%%%

%\begin{figure}
%\centerline{\includegraphics[width=5cm]{fig6}}
%\caption{LexDFS does not number the connected components of $G(V \setminus N[\alpha(1)])$ one after the other.}
%\label{figLexDFScont2}
%\end{figure} 

%%%%%%%%%%%%%%%%%%%%%%%%%%%%%%%

It is shown in \cite{Extremities} that for each labeling structure ${\cal L}$ with a total order on labels and each ${\cal L}$-MLS 
ordering $\alpha$ of a graph $G$, $\alpha(1)$ is an OCF-vertex of $G$, \textit{i.e.} satisfies the property: for each pair $\{y,z\}$ of non-adjacent 
vertices in $N(\alpha (1))$, there is a connected component $C$ of $G(V \setminus N[\alpha(1)])$ such that $y,z \in N_G(C)$. However, the 
definition of a labeling structure given in \cite{Extremities} is less general than the definition given in this paper, as condition IC is 
replaced by conditions (p1): $l \prec Inc(l,i)$ and (p2): $if$ $l \prec l'$ $then$ $Inc(l,i) l \prec Inc(l',i)$: ((p1) and (p2) implies IC 
and IC implies (p1), but does not imply (p2). It turns out that the property of $\alpha(1)$ being an OCF-vertex does not extend to a 
labeling structure defined with condition IC instead of (p1) and (p2) (a counterexample can be built with some effort). It is the only result 
from \cite{Extremities} that does not extend to a labeling structure defined with condition IC.

\section{Conclusion}

In this paper, we explain how a clique tree of a chordal graph $H$ can be computed from an arbitrary peo of $H$, from a pmo of $H$ and from a modification 
of algorithm MLS. We show that a pmo allows to build the cliques of a clique tree one after the other. We characterize labeling 
structures for which it is possible to detect the beginning of a new clique using the labels, and show that each labeling structure can detect this 
with labels when building a clique tree of the complement graph.
\par
Some results concerning algorithm MLS in Section 3 and MLSM in Section 5 are generalizations of results already known for MCS, LexBFS or MCS-M. 
The proofs in this paper largely use Inclusion Condition IC (through Lemma~\ref{lemIC}) and make it clear that condition IC is fundamental for 
the properties of the computed orderings. We believe that many results proved for some particular instances of MLS or MLSM can be proved in a 
more general way for MLS or MLSM using condition IC. This has already been done in \cite{Extremities,MNSM, AtomTree}. We leave open the question of which 
other results could be generalized using condition IC, and which applications could be done of these generalizations.

%%%%%%%%%%%%%%%%%%%%%%%%%%%%%

\end{document}